\documentclass{aa}
\usepackage{natbib}
\usepackage[varg]{txfonts}
\usepackage[colorlinks,linkcolor={blue},citecolor={blue},urlcolor={blue},debug]{hyperref}
\usepackage{epsfig} \usepackage{graphicx}
\usepackage{bbm}
\usepackage{float}
\usepackage[normalem]{ulem}
\usepackage{lipsum}
\usepackage{bm}
\usepackage{scrextend}
\usepackage{breakcites}
\usepackage{amsmath} \usepackage{amsfonts}
\usepackage{amssymb} \usepackage{tikz} \usepackage{tkz-euclide}
\usepackage{comment}
\usepackage[noabbrev]{cleveref}

\usepackage{array,multirow}
\usepackage{footnote}

\makeatletter
\newcommand\footnoteref[1]{\protected@xdef\@thefnmark{\ref{#1}}\@footnotemark}
\makeatother

\providecommand{\DIFdel}[1]{}
\newcommand{\hmpc}{\, h^{-1}\text{Mpc}}
\newcommand{\hsolar}{\, h^{-1} \text{M}_\odot}
\newcommand{\surfmass}{\, h \text{M}_\odot {\rm pc}^{-2}}
\newcommand{\utok}{{ugriZY\!J\!H\!K_{\rm s}}}

\begin{document}

\title{A gravitational lensing detection of filamentary structures connecting luminous red galaxies}

\titlerunning{Lensing detection of filamentary structures}

\author{Qianli Xia\inst{1}\thanks{qx211@roe.ac.uk}
\and Naomi Robertson\inst{2}\thanks{naomi.robertson@physics.ox.ac.uk}
\and Catherine Heymans\inst{1,3} 
\and Alexandra Amon\inst{1,4}
\and Marika Asgari\inst{1}
\and Yan-Chuan Cai\inst{1}
\and Thomas Erben\inst{5}
\and Joachim Harnois-D\'{e}raps\inst{1}
\and Hendrik Hildebrandt\inst{3}
\and Arun Kannawadi\inst{6}
\and Konrad Kuijken\inst{6}
\and Peter Schneider\inst{5}
\and Crist\'{o}bal Sif\'{o}n\inst{7}
\and Tilman Tr\"oster\inst{1}
\and Angus H. Wright\inst{3}
}
\authorrunning{Xia, Robertson et al.}
\institute{
Institute for Astronomy, University of Edinburgh, Royal Observatory, Blackford Hill, Edinburgh EH9 3HJ, UK \and 
Department of Astrophysics, University of Oxford, Keble Road, Oxford OX1 3RH, UK \and
German Centre for Cosmological Lensing, Astronomisches Institut, Ruhr-Universit\"at Bochum, Universit\"atsstr, Bochum, Germany \and
Kavli Institute for Particle Astrophysics \& Cosmology, P.O.Box 2450, Stanford University, Stanford, CA 94305, USA \and
Argelander-Institut f\"ur Astronomie, Universit\"t Bonn, Auf dem H\"ugel 71, 53121, Bonn, Germany \and
Leiden Observatory, Leiden University, P.O.Box 9513, 2300RA Leiden, The Netherlands \and
Instituto de F\'isica, Pontificia Universidad Cat\'olica de Valpara\'iso, Casilla 4059, Valpara\'iso, Chile
}

\date{Accepted 201X. Received 201X ; in original form 201X}

\label{firstpage}

\abstract{We present a weak lensing detection of filamentary structures in the cosmic web, combining data from the Kilo-Degree Survey, the Red Cluster Sequence Lensing Survey and the Canada-France-Hawaii Telescope Lensing Survey.  The line connecting luminous red galaxies with a separation of  $3 - 5\, \hmpc$ is chosen as a proxy for the location of filaments.  We measure the average weak lensing shear around $\sim$11,000 candidate filaments selected in this way from the Sloan Digital Sky Survey.  After nulling the shear induced by the dark matter haloes around each galaxy, we report a $3.4\,\sigma$  detection of an anisotropic shear signal from the matter that connects them.  Adopting a filament density profile, motivated from $N$-body simulations, the average density at the centre of these filamentary structures is found to be $15 \pm 4$ times the critical density.}

\keywords{dark matter, weak gravitational lensing, large-scale structure of the Universe}

\maketitle

\section{Introduction}
\label{intro}

Galaxy surveys, including the 2dF Galaxy Redshift Survey \citep{colless/etal:2001} and the Sloan Digital Sky Survey~\citep[SDSS;][]{zehavi/etal:2011}, have shown that visible matter in our Universe is not uniformly distributed on intermediate scales $\sim 100 \hmpc$.
Instead, a web--like structure is observed with clusters of galaxies identifying the densest regions.
$N$-body simulations predict the existence of these large-scale structures \citep[e.g.,][]{bond/kofman/pogosyan:1996,springel/etal:2005}, suggesting a hierarchical structure formation for the cosmic web. We can classify the web \citep[e.g.,][]{eardley/etal:2015} into regions of  clusters, filaments, sheets and voids. In this cosmic web, large under-dense regions (voids) are enclosed by anisotropically collapsed surface structures (sheets) and line structures (filaments) which intersect at the most over-dense isotropic regions (clusters).
The Zel'dovich approximation predicts that $\sim 42\%$ of the mass of the Universe is in a filament environment \citep{Zel'dovich:1970}, and this has been confirmed by simulations \citep{aragoncalvo/van/jones:2010}. 
However, as filament environments do not display a very high density contrast, this makes direct observations challenging.

One way to observe filaments is from the X--ray emission induced by the warm hot intergalactic medium (WHIM) with several inter-cluster filaments investigated in this way \citep{briel/henry:1995,kull/bohringer:1999,werner/etal:2008}.
There are also reported detections of filaments using overdensities of galaxies \citep{pimbblet/drinkwater:2004, ebeling/barrett/donovan:2004}.
Recently, two independent studies \citep{deGraaff/etal:prep,tanimura/etal:prep} detected the Sunyaev-Zel'dovich (SZ) signal from the ionised gas in the cosmic web.
They estimated the density of ionised gas to be $\sim (28 \pm 12)\%$ of the total baryon density in the Universe, close to resolving the missing baryon problem \citep{bregman:2007}.

In this paper, we investigate the use of weak gravitational lensing to detect filaments.
Based on the distortion of light rays around massive objects, gravitational lensing probes the total mass traced by the large-scale structures and is therefore highly complementary to the SZ detection of the gas.
{ Though \citet{dietrich/etal:2012} made a direct weak lensing detection of a filament connecting two massive Abell clusters,}
the direct detection of typical individual filaments is limited by the low signal-to-noise measurement, and studies instead stack large samples of candidate filaments and analyse the resulting average weak lensing signal \citep{mead/king/mccarthy:2010, clampitt/etal:2016, epps/hudson:2017, kondo/etal:2019}.

\citet[][hereafter C16]{clampitt/etal:2016} determined the weak lensing signal around 135,000 pairs of SDSS Luminous Red Galaxies (LRGs) with a projected separation $6 \hmpc \leq R_{\rm sep} \leq 14 \hmpc$ and a redshift separation of $\Delta z < 0.004$.
Using a `nulling' estimator that cancels the spherically symmetric contribution of the LRG haloes in the shear measurement, they reported a $4.5\sigma$ detection of the filament lensing signal.
In another study, \citet{epps/hudson:2017} used { $\sim 23,000$} pairs of LRGs from the Baryon Oscillation Spectroscopic Survey (BOSS) `LOWZ' and `CMASS' samples as tracers of filaments.
Using data from the Canada–France–Hawaii Telescope Lensing Survey (CFHTLenS), they performed a mass reconstruction of a set of stacked LRG pairs with a projected angular separation between $6 - 10 h^{-1}{\rm Mpc}$ and a redshift separation $\Delta z <0.003$.
After subtracting the signal from a mass reconstruction of a set of stacked LRG pairs with the same separation on the sky, but a greater redshift separation ($0.033< \Delta z < 0.04$) such that haloes should not be physically connected, they reported a $5\sigma$ detection of a filament lensing signal.
{ 
A more recent study \citep{kondo/etal:2019} used { $70,210$} pairs of LRGs from the CMASS sample with a projected separation between $6-14\hmpc$ and a line-of-sight separation of less than $6\hmpc$. 
Using the Subaru Hyper Suprime-Cam (HSC) first-year galaxy shape catalogue and adopting the C16 nulling approach they reported $3.9\,\sigma$ detection of a filament signal.}

We note that the methodology taken in these previous studies can be understood as a three-point galaxy-galaxy-shear correlation function conditioned on specific intervals of separation between lens galaxies \citep{schneider/watts:2005}. The full suite of the galaxy-galaxy-galaxy lensing (GGGL) statistics have been applied to the Red-Sequence Cluster Survey \citep{simon/etal:2008} and CFHTLenS \citep{simon/etal:2019} to measure the excess mass around galaxy pairs separated by $\lesssim$ 300$\,h^{-1} {\rm kpc}$.

In this paper we present the weak lensing signal measured between { $11,706$} LOWZ LRG pairs that have a separation of $3 - 5\, \hmpc$, combining three public weak lensing surveys; the KiDS+VIKING-450 survey \citep[KV450;][]{hildebrandt/etal:2018,wright/etal:2018}, the Red Cluster Sequence Lensing Survey \citep[RCSLenS;][]{hildebrandt/etal:2016} and the CFHTLenS \citep[][]{heymans/etal:2012}.
We improve the nulling methodology described in C16 to deal with contamination from filament tracers and use a large suite of $N$-body simulations to validate our pipeline and compare our results.
A standard \(\Lambda\)CDM cosmology has been adopted throughout this study to calculate distances with a matter density \(\Omega_{\rm m}=0.3\), energy density \(\Omega_{\Lambda}=0.7\), effective number of neutrino species $N_{\rm eff} = 3.04$, baryon density \(\Omega_{\rm b}=0.0\) and current Hubble constant \(H_0=100\,h\,\rm km\,s^{-1}\,Mpc^{-1}\) where \(h\) is the Hubble parameter \(h=0.7\).

This paper is structured as follows.
In Sect.~\ref{Survey}, we describe the survey data and simulations.
Sect.~\ref{form} summarises the weak lensing formalism, the adopted filament model, and methodology.
We show our results in Sect.~\ref{sec:results} and conclude in Sect.~\ref{discuss}.
In Appendix~\ref{app:nulling}, we present a validation of our nulling technique.
In Appendix~\ref{app:sph_rot}, we document the spherical rotation methodology that is required for high declination surveys.
 \section{Surveys and Simulations}
\label{Survey}

\subsection{The Lensing Surveys}
\label{sec:lensing_surveys}

The properties of the three lensing surveys, KV450, RCSLenS and CFHTLenS, are listed in Table~\ref{tab:Survey_details}. They share a similar data processing pipeline, where the shape measurement of galaxies was conducted using the {\emph{lens}}fit model fitting code \citep{miller/etal:2013}.
This approach convolves the pixelised model Point-Spread-Function with an analytical surface brightness model consisting of bulge and disk components. It uses model fitting to estimate galaxy ellipticitices $\epsilon_1^{\rm obs}$ and $\epsilon_2^{\rm obs}$ with an associated inverse variance weight, $w_{\rm s}$. 
The (reduced) shear (c.f. Eq.~\ref{g_pot}) is then given by the weighted average of ellipticities, $\gamma^{\rm obs}_i \approx \sum_{\rm s} w_{\rm s} \epsilon_{i}^{\rm obs} / \sum_{\rm s} w_{\rm s} \; (i=1,2)$.
The observed shear is biased with respect to the true shear and is typically described by the linear bias model \citep{heymans/etal:2006} as
\begin{align}
    \gamma^{\rm obs} = (1+m)\gamma^{\rm true} + c \,,
\end{align}
In all of these three surveys, the shear multiplicative bias terms were characterised as a function of the signal-to-noise ratio and size of the galaxies, thereby allowing us to calculate the bias for an arbitrary selection of galaxies. The correction for this multiplicative bias is carried out as it was in \citet{velander/etal:2014}.

Photometric redshifts, $z_{\rm B}$, were estimated using the Bayesian photometric redshift algorithm \citep[{\sc bpz};][]{benitez:2000} as detailed in \citet{hildebrandt/etal:2012}. 
\citet{wright/etal:2018} and \citet{hildebrandt/etal:2018} show how the photometric redshifts distributions for KV450 are then calibrated using the `weighted direct calibration' method, with weights estimated using a deep spectroscopic training sample in  9-band $\utok$ magnitude space.
No such calibration was performed for the 4-band (RCSLenS) or 5-band (CFHTLenS) surveys, and instead a probability distribution of true redshifts was estimated from the sum of the BPZ redshift probability distributions. 
This approach has been demonstrated to carry more systematic error \citep{choi/etal:2016,hildebrandt/etal:2017}.
We discuss how we take this redshift uncertainty into account in our final analysis in Sect.~\ref{sec:nz_marginalise}. 

\subsection{The BOSS Survey}

We use the Baryon Oscillation Spectroscopic Survey (BOSS) galaxies from the 12th SDSS Data Release \citep{alam/etal:2015} to define candidate filaments. 
Among all three lensing surveys RCSLenS has the most SDSS overlap with almost double that of KV450 or CFHTLenS. Once a robust photometric redshift selection has been applied, however, RCSLenS has only 20\%/30\% the lensing source density in comparison to CFHTLenS/KV450, (see Table~\ref{tab:Survey_details} for details). 

\begin{table*}[h]
\centering
    \begin{tabular}{lccc}
    \hline
                                       &  KV450                   & RCSLenS                 & CFHTLenS                \\
    \hline
    Total area (deg$^2$)               &  454                     & 785                     & 154                     \\
    Unmasked area (deg$^2$)            &  341.3                   & 571.7                   & 146.5                   \\
    Total LOWZ overlap area (deg$^2$)  &  135.91                  & 224.63                  & 113.83                  \\
    $z_{\rm B}$ selection              &  $0.1 < z_{\rm B} < 1.2$ & $0.4 < z_{\rm B} < 1.1$ & $0.2 < z_{\rm B} < 1.3$ \\
    $n_{\rm eff}$ (arcmin$^{-2}$)      &  6.93                    & 2.2                     & 11                      \\
    photometric bands                  &  ($u,g,r,i,Z,Y,J,H,K_{\rm s}$)   & ($g,r,i,z$)             & ($u^*,g',r',i',z'$)     \\
    \hline
    \end{tabular}
    \caption{This table summarises the properties of each of the three lensing surveys used in this analysis; the total and effective unmasked survey area, the total LOWZ overlap area, photometric redshfit, $z_{\rm B}$, selection and the effective number density of lensing sources $n_{\rm eff}$ under the corresponding $z_{\rm B}$ selection. For the $z_{\rm B}$ selection, we followed \citet{hildebrandt/etal:2017}, \citet{hildebrandt/etal:2016} and \citet{heymans/etal:2012} respectively.}
    \label{tab:Survey_details}
\end{table*}

Given the depth of the lensing surveys and the uncertainty in the high redshift tail of the redshift distribution for CFHTLenS and RCSLenS, we choose to limit our analysis to the LOWZ sample, selected based on colour and magnitude, using a redshift cut $0.15 < z < 0.43$ \citep{ross/etal:2012}. We do not consider the higher redshift CMASS sample.
The typical virial halo mass of LOWZ galaxies is $\sim 5.2\times10^{13}\hsolar$ \citep{parejko/etal:2013}. 
These haloes have a typical virial radius $\sim 1 \hmpc$.

\subsection{Simulations}
\label{sims}

We use the Scinet Light Cone Simulations \citep[SLICS\footnote{\url{http://slics.roe.ac.uk}};][]{harnoisderaps/van:2015,harnoisderaps/etal:2018} to test our methodology.
This suite provides us with 819 independent light cones on a 10 $\times$ 10 deg$^2$ patch of the sky. 
Each light cone is constructed from the full non-linear evolution of $1536^3$ particles with $m_p = 2.88 \times 10^9 \hsolar$, within a $505^3(\hmpc)^3$ cube. Particles are then projected onto mass sheets at 18 redshifts between $0 < z < 3$, and subsequently inspected to identify dark matter haloes.
For each simulation, 100 deg$^2$ light cone mass sheets and haloes are extracted; the former are then ray-traced into lensing shear maps, while the latter are used to generate mock LOWZ galaxies with a halo occupation distribution, { that is optimised such that the clustering of mock galaxies is consistent with the LOWZ data, }(see \citealt{harnoisderaps/etal:2018} for details).
{ We find the mock sample variance to be in good agreement with the Jack-knife errors measured directly from the data.}
{
Source galaxy positions are drawn at random, with the shear assigned for a range of redshifts and high number density.  We next randomly draw from the SLICS mock source galaxy sample so as to match the corresponding number density and redshift distribution of each of the three surveys, KV450, RCSLenS and CFHTLenS.  Intrinsic galaxy shapes are chosen to match the KiDS ellipticity dispersion \citep{hildebrandt/etal:2017}, which is good description of the ellipticity dispersion also found in RCSLenS and CFHTLenS.
}
 \section{Summary of Weak Lensing Formalism and Methodology}
\label{form}

In this section we summarise weak gravitational lensing theory, following the more detailed derivations in \cite{bartelmann/schneider:2001}. 
Assuming the thin lens approximation, a foreground object at a position $\bm{\theta}$ has a 2D comoving surface mass density $\Sigma(\bm{\theta})$. 
The convergence is then defined as
\begin{equation}
\kappa(\bm{\theta})=\frac{\Sigma(\bm{\theta})}{\Sigma_{\rm{crit}} } \, ,
\label{kappa}
\end{equation}
where $\Sigma_{\rm{crit}}$ is the comoving critical surface density in a flat Universe given by
\begin{equation}
\Sigma_{\rm{crit}} = \frac{c^2}{4\pi G}\frac{\chi(z_{\rm s})}{[\chi(z_{\rm s})-\chi(z_{\rm l})]\chi(z_{\rm l})(1+z_{\rm l})} \, .
\label{sigma_crit}
\end{equation}
Here $\chi$ is the comoving distance and $z_{\rm l}, z_{\rm s}$ are the redshifts of the lens and source respectively. 
Since we are interested in the large scale comoving surface density around filaments, this is the appropriate choice of $\Sigma_{\rm{crit}}$ (see \citealt{dvornik/etal:2018} for a discussion on the different definitions of $\Sigma_{\rm{crit}}$).

The deflection potential, $\psi (\theta)$, is connected to the convergence, $\kappa$, via Poisson's equation \( \nabla^2 \psi = 2\kappa\), and the complex shear is related to the second derivatives of the deflection potential via
\begin{equation}
\gamma = \gamma_1 +i\gamma_2 =\frac{1}{2}\left(\frac{\partial ^2\psi}{\partial x_1^2} - \frac{\partial^2\psi}{\partial x_2^2}\right)+i\frac{\partial^2 \psi}{\partial x_1\partial x_2} \, ,
\label{g_pot}
\end{equation}
where $x_1, x_2$ are the horizontal and vertical displacements on the projected sky.

For a filament aligned with the $x_1$ axis if we assume the deflection potential $\psi$ and convergence $\kappa$ are both invariant along the filament, then Eq.~
\ref{g_pot} immediately implies that partial derivatives with respect to the $x_1$-axis will be equal to zero. 
This leads to the approximation that for filaments, we should expect to measure $\gamma_1 \approx - \kappa $  and $\gamma_2 \approx 0$. 
Motivated by the simulation results of \citet{colberg/krughoff/connolly:2005}, \citet{mead/king/mccarthy:2010} considered the power law density profile around filaments and suggested the model for the convergence at a distance $r$ from filament centre, measured perpendicular to the major filament axis ($x_1$-axis)
\begin{align}
    \kappa(r) \approx \frac{\kappa_c}{1 + \left(\frac{r}{r_c}\right)^2}.
    \label{eqn:fitting}
\end{align}
Here $\kappa_c$ is the amplitude of the convergence at the filament centre ($r=0$) and $r_c$ is the half-maximum radius of the density profile.

\subsection{Filament candidiate}

\citet{colberg/krughoff/connolly:2005} showed that cluster pairs separated by $< 5\hmpc$ are always connected by filamentary structures. 
We therefore select Luminous Red Galaxy (LRG) pairs in the LOWZ catalogue with redshift separation $\delta z < 0.002$ and a projected separation $3 \hmpc \leq R_{\rm sep} \leq 5 \hmpc$ as our candidate filaments\footnote{The average 3D separation between these filament candidates is about $7\hmpc$.}  which we will refer to as our physical pairs (PP).
Non-physical pairs (NP) are defined to have the same projected separation range 
but with large line-of-sight separations with $0.033 < \delta z < 0.04$ (corresponding to $\sim 100 \hmpc$).  With such a large physical separation we would not expect to detect a filament signal.  The NP therefore provide an important null-test for our methodology.

Our candidates differ from the selection made by C16, \citet{epps/hudson:2017,deGraaff/etal:prep,tanimura/etal:prep} and { \citet{kondo/etal:2019}}, who focused on separations of $6 - 10 \hmpc$. Our choice maximises signal-to-noise, as shown in our analysis of numerical simulations in Sect.~\ref{sec:simulations}, but for completeness we also present an analysis of $6 -10 \hmpc$ filaments in Sect.~\ref{discuss}

\subsection{Stacking Method}
\label{Method}
For each lens filament candidate at redshift $z_{\rm l} = z_{\rm f} = (z_{\rm lens 1} + z_{\rm lens 2})/2$ we measure the $\Sigma_{\rm crit}$-weighted shear, $E_{\rm f}$, on a grid $(i,j)$ centred and oriented with the pair of LRGs, where
\begin{align}
    E_{\rm f}(i,j) = \sum\limits_{\rm s} w_{\rm s}\,\overline{\Sigma^{-1}_{{\rm crit}} }(z_{\rm l})\,
    \widetilde{\epsilon}^{\rm obs}_{{\rm s}}\,\Theta_{\rm s}(i,j),
    \label{eqn:weightedshear}
\end{align}
and the sum is taken over all sources\footnote{We use a redshift cut $z_{\rm B} - z_{\rm l} > 0.1$ everywhere to ensure that the majority of our source galaxies are behind the foreground galaxies, and not associated with them. 
Fig.~D3 in \citet{amon/etal:2018} demonstrates that with this selection, contamination of the KiDS source sample is negligible for the scales we probe at $> 0.1 \hmpc$.}, s, with $z_{\rm B} > z_{\rm l} + 0.1$ and
\begin{align}
    \Theta_{\rm s}(i, j) &= \left\{
            \begin{array}{lll}
              1 & \quad \text{if source, s, lies in pixel }  (i,j),\\
              0 & \quad \text{otherwise}.
            \end{array}\right.
\end{align}
The complex ellipticity $\widetilde{\epsilon}^{\rm obs}_{{\rm s}} = \widetilde{\epsilon}^{\rm obs}_{{\rm 1}} + i\widetilde{\epsilon}^{\rm obs}_{{\rm 2}}$ is the observed ellipticity of the source rotated into the reference frame where the filament lies along the $x_1$ axis.
The $\Sigma_{\rm crit}$ weight converts from a shear estimate to an estimate of the surface mass density $\Sigma$ in Eq.~\ref{kappa}.
The grid $(i,j)$ has an extent $[-2R_{\rm sep}, 2R_{\rm sep}]\times[-2R_{\rm sep}, 2R_{\rm sep}]$ and $129^2$ pixels, and the pair of lens galaxies that define the filament candidate are positioned to lie at the centre of the pixels at $(-0.5 R_{\rm sep}, 0)$ and $(0.5 R_{\rm sep},0)$.

We also construct a corresponding weight map for each filament candidate
\begin{align}
    W_{\rm f}(i,j) = \sum\limits_{\rm s} w_{\rm s} \, \left[\overline{\Sigma^{-1}_{{\rm crit}} }(z_{\rm l})\right]^2
    \,\Theta_{\rm s}(i,j),
    \label{eqn:weights}
\end{align}
where the extra factor of $\overline{\Sigma^{-1}_{\rm crit}} (z_{\rm l})$  provides optimal signal-to-noise weighting \citep{velander/kuijken/schrabback:2011}.

When rotating each filament pair into a common reference frame, we note that at high declination, the tangent plane method used in \citet{epps/hudson:2017} and the direct cartesian approximation in C16 results in non-uniform grid cells. 
As the nulling approach requires a flat geometry on the grid, we found that these approximations lead to a biased result.
To solve this problem for high-declination patches, we use the spherical rotation method from \citet{deGraaff/etal:prep}. 
This process is detailed in Appendix~\ref{app:sph_rot}, and illustrated in Fig.~\ref{fig:sph_rot}, with the rotated shear map $(\widetilde{\epsilon_1}, \widetilde{\epsilon_2})$ defined in Eq.~\ref{e_rotation}.

\begin{figure}[h]
    \centering
    \includegraphics[width=0.95\columnwidth]{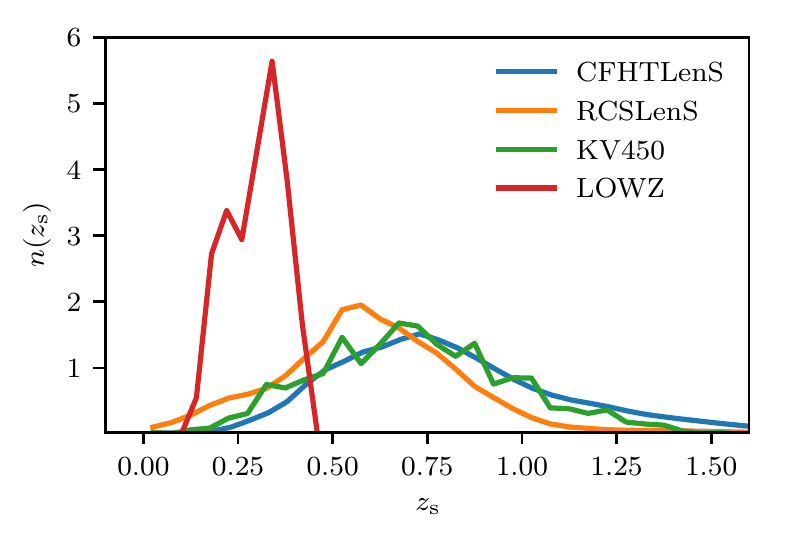}
    \caption{The effective $n(z)$ of all three lensing surveys, as defined in Eq.~\ref{eqn:nofz}, when using LOWZ galaxies as lenses. Each curve is normalised such that $\int n(z)\,\mathrm{d}{z} = 1$.}
    \label{fig:nz-plot}
\end{figure}

As we have spectroscopic redshifts for the filaments but only photometric redshifts for the sources, the inverse critical surface mass density $\Sigma_{\rm crit}^{-1}(z_{\rm l})$ is calculated for each survey as 
\begin{align}
    \overline{\Sigma^{-1}_{\rm{crit}} }(z_{\rm l}) & \equiv \int_{z_{\rm l}}^{\infty} \!\mathrm{d}z_{\rm s} \, 
    \overline{p_{\rm s}}(z_{\rm s}, z_{\rm l}) \Sigma_{\rm{crit}}^{-1} (z_{\rm l},z_{\rm s}) \\
     & = \frac{4 \pi G(1+z_{\rm l})\chi(z_{\rm l})}{c^2} \int_{z_{\rm l}}^{\infty} \!\mathrm{d}z_{\rm s} \, \overline{p_{\rm s}}(z_{\rm s}, z_{\rm l}) \left[1- \frac{\chi(z_{\rm l})}{\chi(z_{\rm s})} \right], 
     \label{eqn:sig_inv}
\end{align}
where $\overline{p_{\rm s}}$ is the probability distribution of the true redshift of the source galaxies that enter the measurement 
\begin{align}
    \overline{p_{\rm s}}(z_{\rm s}, z_{\rm l}) = 
    \frac{\sum\limits_s w_{\rm s} p_{\rm s}(z_{\rm s}|z_{\rm B})}
    {\sum\limits_s w_{\rm s}}.
    \label{eqn:pofz}
\end{align}
For CFHTLenS and RCSLenS we use the per-source $p_{\rm s}(z_{\rm s}|z_{\rm B})$ provided by each survey, even though this has been shown to introduce biases \citep{choi/etal:2016}, which we account for in Sect.~\ref{sec:nz_marginalise}.
For KV450 we use the weighted direct calibration  method of \citet{hildebrandt/etal:2017} to determine the source redshift distribution $\overline{p_{\rm s}}(z_{\rm s})$ directly for an ensemble of sources. 
In practice we calculate $\overline{\Sigma^{-1}_{\rm{crit}} }$ in Eq.~\ref{eqn:sig_inv} at eight $z_{\rm l}$ values and interpolate to evaluate $\overline{\Sigma^{-1}_{\rm{crit}} }$ at each filament redshift.
The effective $n(z_{\rm s})$ for each survey, given by 
\begin{align}
    n(z_{\rm s}) = \int \, \overline{p_{\rm s}}(z_{\rm s}, z_{\rm l}) p(z_{\rm l}) \,\mathrm{d}{z_{\rm l}}, 
    \label{eqn:nofz}
\end{align}
is shown in Fig.~\ref{fig:nz-plot}, with CFHTLenS and KV450 providing a deeper source redshift than RCSLenS.

We correct the measured shear signal $E_{\rm f}$, with the signal measured around `random' filaments. This is now a standard procedure in galaxy-galaxy lensing studies \citep[e.g.,][]{mandelbaum/etal:2005} which removes any hidden systematics and reduces sampling variance noise.
We create random filament samples for each survey patch, listed in Table~\ref{tab:patch}, by randomly shifting filaments within the same patch while preserving their redshifts, position angles and number density in the patch.
\begin{table}[h]
    \centering
    \begin{tabular}{l|lcc}
    \hline
Survey  & Patch Name &  $A^{\rm LOWZ}_{\rm eff}$   &  $N_{\rm fil},3\sim5\hmpc$ \\
    \hline
CFHTLenS&   W1$^{\triangle}$       &      53.38     &    1106   \\
        &   W3$^{\triangle}$       &      40.12     &     835   \\
        &   W4$^{\triangle}$       &      20.34     &     528   \\
    \hline
KV450   &   G9                     &      11.10     &     305   \\
        &   G12$^{\triangle}$      &      30.08     &     586   \\
        &   G15$^{\triangle}$      &      94.73     &    2150   \\
        
    \hline
RCSLenS &   R0047$^{\triangle}$    &      40.27     &    2111   \\
        &   R0133                  &      14.25     &     642   \\
        &   R1040$^{\triangle}$    &      26.94     &     580   \\
        &   R1303                  &      4.00      &     119   \\
        &   R1514$^{\triangle}$    &      32.72     &    1296   \\
        &   R1613                  &      9.16      &     331   \\
        &   R1645$^{\triangle}$    &      22.66     &     678   \\
        &   R2143$^{\triangle}$    &      42.16     &    1063   \\
        &   R2329$^{\triangle}$    &      32.09     &     773   \\
        &   R2338                  &      0.39      &      25   \\
    \hline
    \end{tabular}
    \caption{Table showing the effective area and number of filaments in each survey patch. The patches that are used in the analysis are identified with a $\triangle$.}\label{tab:patch}
\end{table}
As we do not expect any physical signal from the random catalogue, we subtract any measured `random' signal from the data as follows:
\begin{align}
    E_{\rm f}^{\rm cor} = E_{\rm f} - \overline{E_{\rm ran}}.
\end{align}
Here, for a patch with $N_p$ filament candidates $\overline{E_{\rm ran}}$ is given by 
\begin{align}
    \overline{E_{\rm ran}} = \frac1{N_R}\frac1{N_p}\sum\limits_{r = 1}^{N_R}w_{\rm ran}\sum\limits_{k=1}^{N_p}E_r,
\end{align}
and
\begin{align}
w_{\rm ran} = 
         {\sum\limits_{k=1}^{N_p}w_{\rm f}W_{\rm f}} / 
         {\sum\limits_{k=1}^{N_p}W_{\rm r}},
\end{align}
where $E_r$ and $W_r$ are the weighted shear (Eq.~\ref{eqn:weightedshear}) and weight (Eq.~\ref{eqn:weights}) but measured around a random pair. $w_{\rm ran}$ is the normalisation weight 
where $N_R$ is the number of realisations which ensures $N_R\times N_p$ exceeds 100,000 in each patch\footnote{\label{foot:patch}For patches with $A_{\rm eff} < 20\,{\rm deg}^2$  we found that the sampling variance between the random catalogues was too large and we therefore do not use these patches in the final analysis.}. This ensures that the random signal has low scatter so that we can take the mean as the random correction. 

The random-corrected shear map $E_{\rm f}^{\rm cor}$ and weight map $W_{\rm f}$ are then optimally combined over all filament candidates to determine the total weighted shear signal, $\mathcal{T}$, over the full sample, 
\begin{align}
    \mathcal{T}(i,j) = \frac1 K
                  \frac{\sum\limits_{\rm f}w_{\rm f} E_{\rm f}^{\rm cor}}
                       {\sum\limits_{\rm f}w_{\rm f} W_{\rm f}}
\end{align}
where $ w_{\rm f} = w_{\rm lens1} \times w_{\rm lens2}$ is the {product} of the recommended SDSS completeness weights for the LOWZ galaxies { to account for any extra galaxy pairs in the case of fibre collision}.
We have also applied the multiplicative calibration correction at this stage which is given by \citet{velander/etal:2014} as
\begin{align}
    K(i,j) = \frac{\sum\limits_{\rm f}w_{\rm f} 
                  \sum\limits_{\rm s} w_{\rm s}\, 
                  \left[\overline{\Sigma^{-1}_{{\rm crit}} }(z_{\rm f})\right]^2\,
                  (1 + m_{\rm s}) \,
                  \Theta_{\rm s}(i,j)}
                 {\sum\limits_{\rm f}w_{\rm f} W_{\rm f}}.
\end{align}

The total weighted shear $\mathcal{T}(i,j)$ is a combination of both the shear contribution from the haloes surrounding the LOWZ LRGs and the contribution from any filament that connects them.

In order to isolate the filament we apply the ``nulling'' procedure described in Appendix~\ref{app:nulling} to get a final measurement of the shear contribution from the filament only, $\mathcal{F}(r)$, as a function of the distance, $r$, from the central filament axis,
\begin{align}
    \mathcal{F}(r) = \frac{\sum\limits_{i=i_{\rm min}}^{i_{\rm max}} \mathcal{N} [\mathcal{T}(i, r)]}
                          {i_{\rm max} - i_{\rm min} + 1}.
    \label{eqn:measurement}
\end{align}
Here $\mathcal{N}$ is the nulling operator given in Eq.~\ref{eqn:nulling} and the summation over pixels from an $i_{\rm min}$ to $i_{\rm max}$ runs along the filament from $-0.438 R_{\rm sep}$ to $0.438 R_{\rm sep}$ . This value was found to minimise any residual contribution from the haloes positioned at $\pm 0.5 R_{\rm sep}$, that remains after a nulling analysis of the SLICS simulation (see Sect~\ref{sec:simulations}).
Our nulling operator $\mathcal{N}$ combines the shear values measured at 8 different positions {(including 4 positions from a reflection about the filament axis)} which alternatingly rotate  around the two haloes. 
In this way the isotropic contribution from the parent haloes sum to zero (i.e., ``null'') and any anisotropic contribution in-between the two haloes can be recovered. 
In Appendix~\ref{app:nulling} we provide a detailed proof and compare our nulling approach to that adopted in C16. Through tests on a fiducial model we show that the C16 nulling approach produces a biased result on large scales.

In this derivation we have carried both components of the shear with $\mathcal{F} = \mathcal{F}_{\gamma_1} + i \mathcal{F}_{\gamma_2}$.
Given the \citet{mead/king/mccarthy:2010} filament model, where $\kappa = -\gamma_1$, we expect $\mathcal{F}_{\gamma_2} = 0$. We will fit $\mathcal{F}_{\gamma_1}$ using the two-parameter model in Eq.~\ref{eqn:fitting} with the amplitude parameter replaced by $\mathcal{F}_c$ which is equivalent to $\Sigma_{\rm crit} \kappa_c$ for a single lens-source pair.

\subsection{Error Estimation {from SLICS}}
\label{sec:boot}
In order to estimate the error on the measured signal from observations, { we use a large number of independent and representative lensing simulations from SLICS for each of the three surveys.
SLICS allows us to correctly account for the sampling variance, which was found to be the dominant source of noise in \citet{kondo/etal:2019}.
The source sample of galaxies differs for each filament pair owing to our source selection that $z_B - z_{\rm f} > 0.1$. For KV450 we can apply this source selection criteria accurately as the SLICS simulations include mock KV450 photometric redshifts that re-produce the scatter, bias and catastrophic outlier populations found in the KV450 data \citep{harnoisderaps/etal:2018}.
For CFHTLenS and RCSLenS, this information is not encoded. We therefore create mocks from SLICS, modelling source samples for four different filament bins with $(z_{\rm min}, z_{\rm max}) = [0.1,0.2], [0.2,0.3], [0.3,0.4]$ and $[0.4,0.5]$ respectively. For each filament bin we calculate the source galaxy redshift distribution $n(z)$, using Eq.~\ref{eqn:nofz} and the effective galaxy number density for sources with $z_B - z_{\rm max} > 0.1$. We then populate 500 independent simulations using these distributions, and measure and combine the weighted shear and weight maps for each of the four filament bins. For KV450 we are able to verify that this binned approach is consistent to the unbinned methodology applied to the data using the KV450 SLICS simulations.

The covariance matrix of the signal is reweighted by the number of filament candidates for each survey, $n_{\rm fil, survey}$, and estimated from the SLICS simulations as 
\begin{align}
  {\rm \bf Cov} =   \frac{\bar{n}_{\rm fil, sim}}{n_{\rm fil, survey}}
                    \frac1{N_{\rm sim}-1}
                    \sum\limits_{k=1}^{N_{\rm sim}}
                            ({\bm{\mathcal{F}}^{k}_{\gamma_{1}} } - 
                                \overline{\bm{\mathcal{F}}_{\gamma_{1}} })
                            ({\bm{\mathcal{F}}^{k}_{\gamma_{1}} } - 
                                \overline{\bm{\mathcal{F}}_{\gamma_{1}} })^T,
    \label{eqn:cov0}
\end{align}
where $\overline{\bm{\mathcal{F}}_{\gamma_{1}}}$ is the filament signal (Eq.~\ref{eqn:measurement}) averaged over all $N_{\rm sim}=500$ survey-specific SLICS simulations, and $\bar{n}_{\rm fil, sim}$ is the average number of filament candidates in these simulations.
The covariance is then used to calculate the $\chi^2$ when estimating parameters in the filament model as
}
\begin{align}
    \chi^2_{\rm model} = 
    (\bm{\mathcal{F}}_{\gamma_{1}} - 
        \bm{\mathcal{F}}_{\gamma_1}^{\rm fit})^T
    {\rm \bf Cov}^{-1}
    (\bm{\mathcal{F}}_{\gamma_{1}} - 
        \bm{\mathcal{F}}_{\gamma_1}^{\rm fit}),
\end{align}
where  $\bm{\mathcal{F}_{\gamma_1}}^{\rm fit} = \bm{\mathcal{F}_{\gamma_{1}} }^{\rm fit}(\mathcal{F}_c,r_c,r)$ is our filament model defined in Eq.~\ref{eqn:fitting}, calculated on a fine grid of parameters $(\mathcal{F}_c, r_c)$.
In analogy to Eq.~\ref{eqn:cov0}, we also define a covariance for the $\mathcal{F}_{\gamma_2}$ component which serves as a systematic null-test for our analysis. 

We found that a simple bootstrap error analysis of the data, where the set of maps are resampled with repetitions before stacking, underestimates the true measurement error. This approach misses the sampling variance term which, like \citet{kondo/etal:2019}, we find is a significant component to the error for small-area surveys such as KV450 and CFHTLenS.

\subsection{Accounting for uncertainty in the redshift distributions}
\label{sec:nz_marginalise}

As discussed in Sect.~\ref{sec:lensing_surveys}, the probability distribution $\overline{p_{\rm s}}(z_{\rm s}, z_{\rm l})$ from Eq.~\ref{eqn:pofz} have not been calibrated for RCSLenS and CFHTLenS. 
A systematic uncertainty on the resulting $n(z_{\rm s})$ is thus expected. 
In order to take this into account, we use a nuisance parameter $\delta z_{\rm s} = 0.1$ for RCSLenS and $\delta z_{\rm s} = 0.04$ for CFHTLenS that captures
 the $p(z_{\rm s})$ uncertainty determined by \citet{choi/etal:2016}. For KV450 we use $\delta z_{\rm s} = 0.025$ following \citet{wright/etal:2018}. 
We shift the $\overline{p_{\rm s}}(z_{\rm s}, z_{\rm l})$ by $\pm \delta z_{\rm s}$ in Eq.~\ref{eqn:sig_inv} to yield two new functions $\overline{\Sigma^{-1}_{\rm{crit}} }(z_{\rm l})^{\pm}$, and repeat the full measurement and error analysis using both the $\overline{\Sigma^{-1}_{\rm{crit}} }(z_{\rm l})^{+}$ and $\overline{\Sigma^{-1}_{\rm{crit}} }(z_{\rm l})^{-}$ calibration. We then estimate the likelihood of the data $D$ given the model $(\mathcal{F}_c, r_c)$, assuming a prior on the uncertainty $\delta z_{\rm s}$ which consists of 3 $\delta$-functions and marginalised over as 
\begin{align}
    \mathcal{L}(D | \mathcal{F}_c, r_c) 
    &=  \mathcal{L}(D | \mathcal{F}_c, r_c, -\delta z_{\rm s}) + 
        \mathcal{L}(D | \mathcal{F}_c, r_c,  0) + 
        \mathcal{L}(D | \mathcal{F}_c, r_c,  \delta z_{\rm s}) \nonumber \\
    &\propto \exp\left(-\frac{1}{2}\chi_{-\delta z_{\rm s}}^2\right) + \exp\left(-\frac{1}{2}\chi^2\right) + \exp\left(-\frac{1}{2}\chi_{\delta z_{\rm s}}^2\right).
    \label{likelihood}
\end{align}
A full marginalisation where many samples are taken at different redshift offsets, spanning the $\delta z$ range, is unfortunately unfeasible given the complexity of the measurement pipeline.
 \begin{figure}[h]
    \centering
\includegraphics[width=0.95\columnwidth]{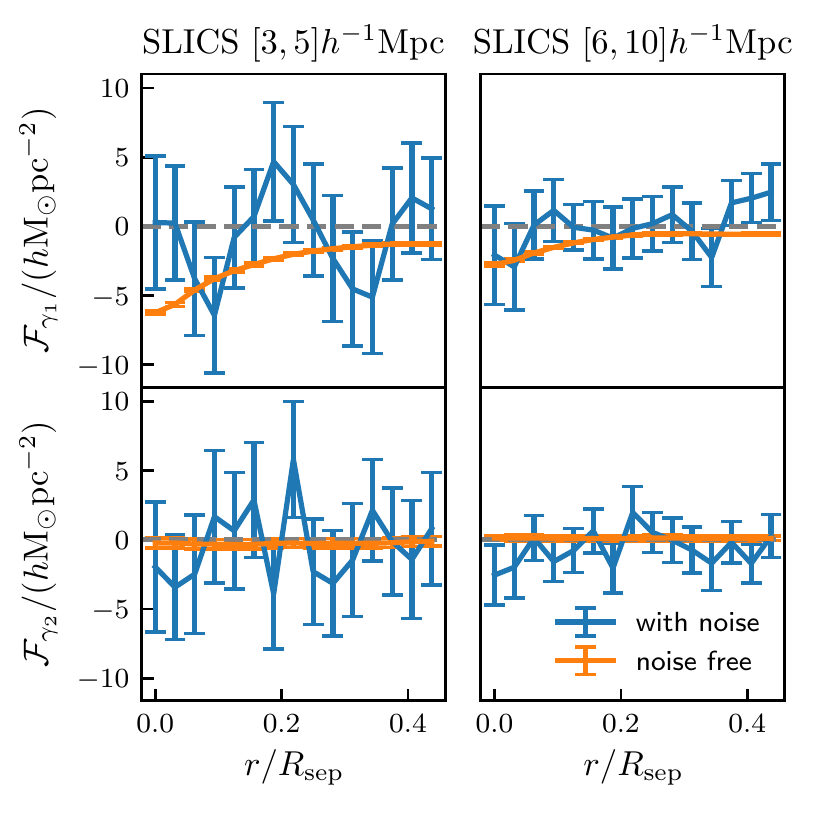}
    \caption{The weighted shear measured between LRG pairs in the SLICS simulation of a random 400 ${\rm deg}^2$ degree survey. \underline{Left:} Results from $3-5\hmpc$ filament candidates. The upper panel shows the average surface mass density $\mathcal{F}_{\gamma_1}$, and the lower panel shows the measured cross-shear component. On both panels, the blue data shows the result for a KiDS-like survey depth and shape noise, and the orange data points show the measurement for a noise-free simulation with the errorbar given by the error on the mean of all 158 realisations. \underline{Right:} The set of results from $6-10\hmpc$ filament candidates.}
    \label{fig:SLICS_Results}
\end{figure}

\section{Results}
\label{sec:results}

\subsection{Filaments in SLICS}
\label{sec:simulations}

We validate our pipeline using the SLICS simulations of mock LOWZ lens galaxies and mock {KiDS} sources. 
\begin{figure*}
    \centering
    \includegraphics[width=\textwidth]{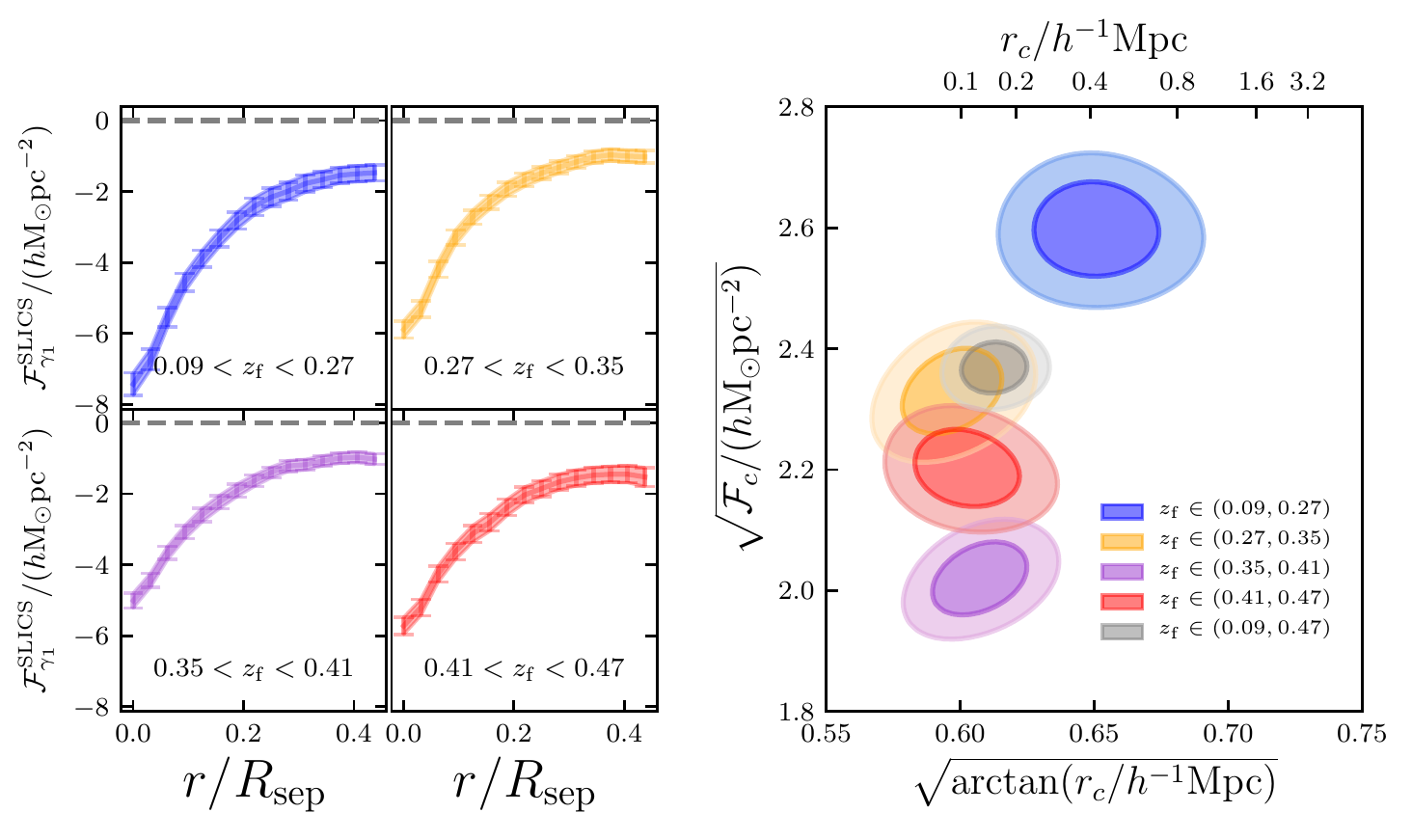}
    \caption{The redshift evolution of the filament signal in the noise-free SLICS simulation. The left panel shows the $\mathcal{F}_{\gamma_1}$ signal measured from filament candidates in SLICS for four redshift quantiles. The right panel shows the 68\% and 95\% confidence region of the model parameters $(\mathcal{F}_c, r_c)$ from the corresponding signal in the left panel, with the result from all samples combined shown in grey.}
    \label{fig:evol}
\end{figure*}
{
We analyse both a noise-free catalogue and a catalogue with shape noise and mock photometric redshifts.
The result is shown in Fig.~\ref{fig:SLICS_Results} where the upper panels show measurements of $\mathcal{F}_{\gamma_1}$ and the lower panels show measurements of $\mathcal{F}_{\gamma_2}$. 
The left and right columns correspond to results from the $3-5 \hmpc$ and $6-10 \hmpc$ filament candidates respectively.
}

For the noise-free simulations, we find that $\mathcal{F}_{\gamma_2}$ is consistent with zero for both the $3-5 \hmpc$ and $6-10 \hmpc$ length filaments. This demonstrates that our nulling procedure correctly removes the contribution from the LRG haloes in the analysis.
However, for the simulations with shape noise, both the $\mathcal{F}_{\gamma_1}$ and $\mathcal{F}_{\gamma_2}$ measurements are consistent with zero, which suggests that even though KV450 is deeper than the KiDS-450 data simulated in SLICS, we should not expect a significant detection from KiDS alone.
As the variance in the noise-free simulation reflects the level of sample variance, we also report that, by measuring the noise level from these two sets of simulations, the sample variance is comparable with the shape noise. 

{
Constraining the parameters of the \citet{mead/king/mccarthy:2010} model with the noise-free SLICS results we find $\mathcal{F}_c^{3-5} = 5.61\pm0.55 \surfmass,~r_c^{3-5} = 0.40 \pm 0.04\hmpc,~\mathcal{F}_c^{6-10} = 2.25\pm0.14\surfmass,$ and $r_c^{6-10} = 1.12 \pm 0.08\hmpc$. The $\chi^2_{3-5} = 16.33$ and $\chi^2_{6-10} = 15.42$ demonstrate that the model is a good fit to the data ($\nu = 13$ degrees of freedom).
We find that the surface mass density of the filament is a factor of 2.5 smaller for the $6- 10 \hmpc$ filament and will therefore be more challenging to detect using gravitational lensing.
}

With the noise-free simulations we are able to analyse whether our signal depends on the redshift of the filament.
We constrain the amplitude $\mathcal{F}_c$ and the scale $r_c$ parameters of the filament model for 4 redshift quantiles of the SLICS LOWZ filament samples using the same background sources, with the result shown in Fig.~\ref{fig:evol}.
The choice of parameterisation on the axes is motivated by the filament model equation as well as for visual simplicity.
Here we see significant differences between the samples which could be caused by an evolution in the bias of the LOWZ-like galaxy samples in the SLICS mocks, and/or the evolution of the filament density field.
Given the different source redshift distributions of the three lensing surveys (see Fig.~\ref{fig:nz-plot}) which makes the effective redshift of the average lens differ, this result suggests that we should not necessarily expect the results of these surveys to agree perfectly.

\subsection{The detection of filaments with KV450, RCSLenS and CFHTLenS}
\label{sec:detection}
\renewcommand{\arraystretch}{1.5}
\begin{table*}
    \centering



{
    \begin{tabular}{c|ccccc}
    \multicolumn{1}{c}{}&\multicolumn{1}{c}{}& CFHTLenS & RCSLenS & KV450 & All \\
\hline
    \multirow{7}{*}{PP}       & $\mathcal{F}_c / (h {\rm M_\odot pc^{-2}})$            & $13.3^{+4.1}_{-4.0}$   & $14.4^{+8.5}_{-7.6}$ & $4.6^{+5.9}_{-4.5}$ & $10.5^{+2.9}_{-2.8}$\\
    &                           $r_c / (\hmpc)$                      & $ 0.5^{+0.4}_{-0.2}$   & $ 0.2^{+0.3}_{-0.1}$ & $-$ & $ 0.4^{+0.2}_{-0.1}$\\

    &                           $\chi^2_{\rm min, model}$                           & 12.2  & 11.6   &  7.5   &  40.4  \\
    \cline{2-6}
    &                           $\chi^2_{{\rm min, null},\mathcal{F}_{\gamma_1}}$   & 24.6  &  15.6  &  8.1   &  55.0  \\
    &                           $\sigma_{\mathcal{F}_{\gamma_1}}$                   &  3.08 &  1.49  &  0.35  &  3.40  \\
    &                           $\chi^2_{{\rm min, null},\mathcal{F}_{\gamma_2}}$   &  6.8  &  6.78  &  2.61  & 22.06  \\
    &                           $\sigma_{\mathcal{F}_{\gamma_2}}$                   &  0.05 &  0.05  &  2e-4  &  2e-3  \\
    \hline
    \multirow{4}{*}{NP}       & $\chi^2_{{\rm min, null},\mathcal{F}_{\gamma_1}}$   &  9.2  &  20.0  &  19.3  &  53.8  \\
    &                           $\sigma_{\mathcal{F}_{\gamma_1}}$                   &  0.17 &  1.38  &  1.28  &  1.36  \\
    &                           $\chi^2_{{\rm min, null},\mathcal{F}_{\gamma_2}}$   &  12.3 &  20.7  &  5.8   &  43.0  \\
    &                           $\sigma_{\mathcal{F}_{\gamma_2}}$                   &  0.44 &  1.45  &  0.02  &  0.59  \\
    \hline
    \end{tabular}
    \caption{$\chi^2$ value and p-value for all computed signals from each individual survey and their combination. For each survey, PP means physical pair and NP stands for non-physical pair. Both NP and $\mathcal{F}_{\gamma_2}$ serve as null tests for our analysis. For the combined signals, the degree of freedom is $\nu = 45 -2$. We also note that $r_c$ is unconstrained by KV450.}
    \label{tab:individual_chisquares}
}
\end{table*}

\begin{figure*}
\centering
    \includegraphics[width=\textwidth]{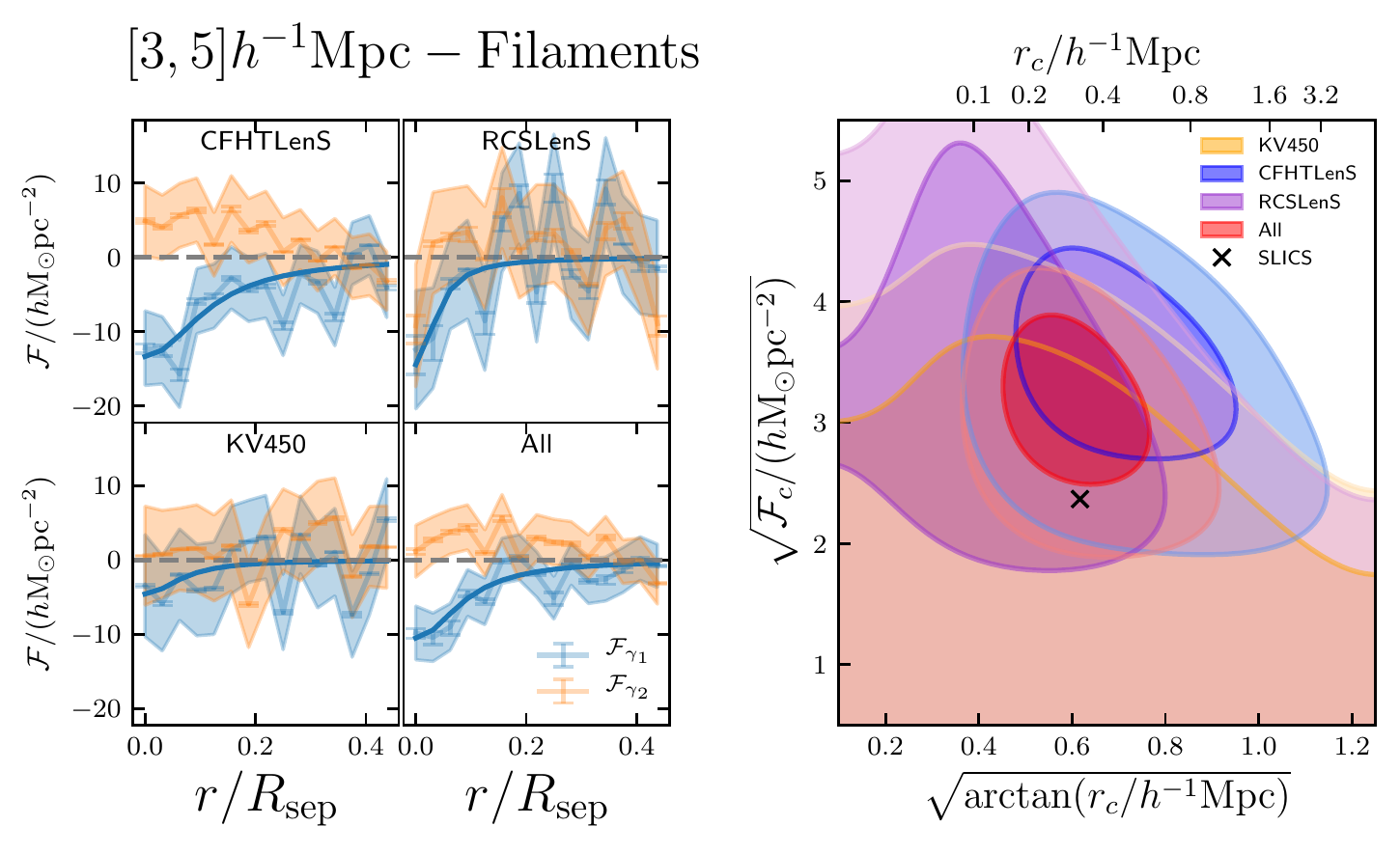}
    \includegraphics[width=\textwidth]{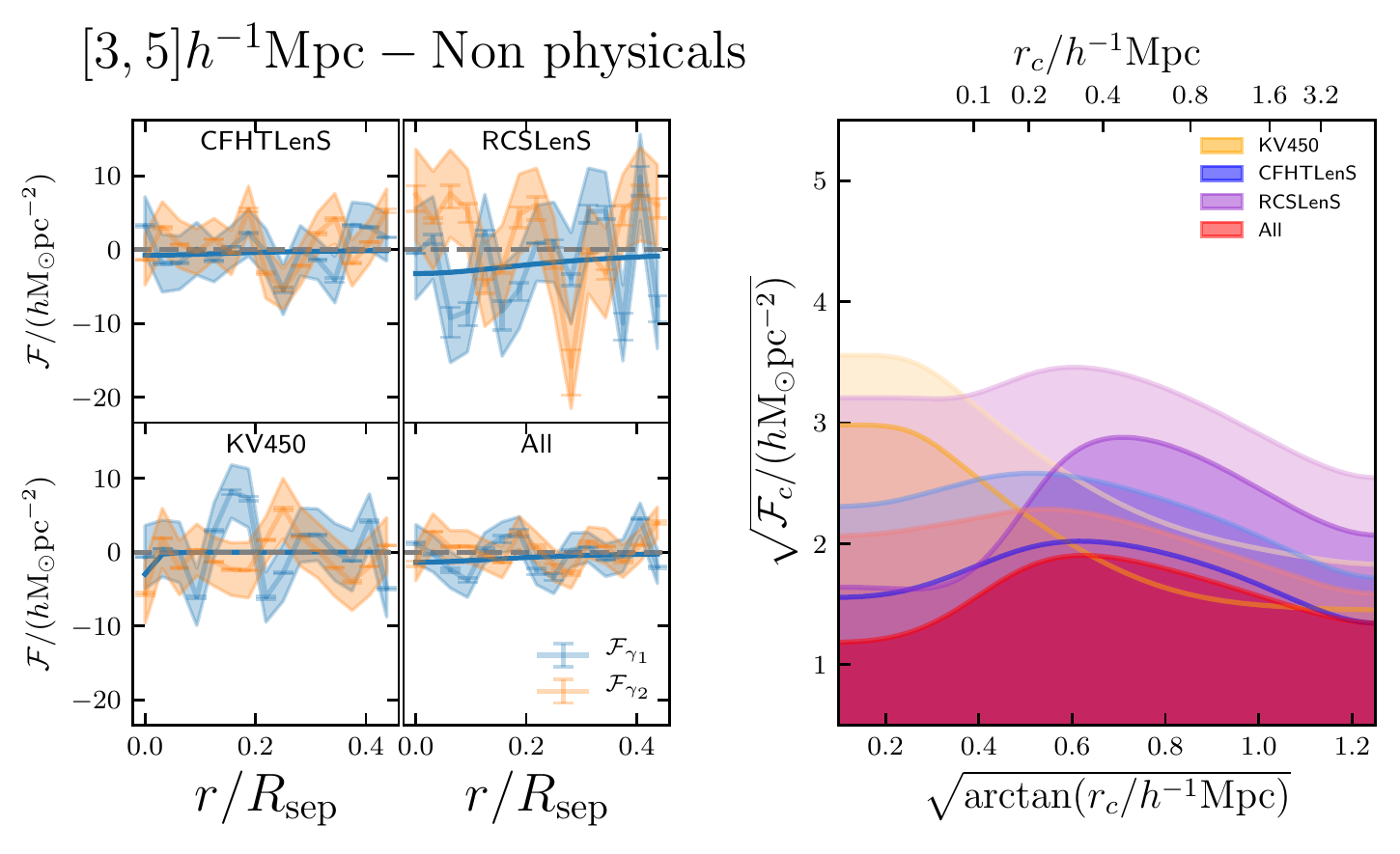}
    \caption{The detection of the cosmic web between neighbouring luminous red galaxies as detected through the weak lensing of background galaxies from different lensing surveys.
    \underline{Left:} The $x$-axis is the distance measured perpendicular to the filament axis scaled such that 1 is equivalent to the projected separation between the pair of LRGs. 
    The $y$-axis is the nulled shear signal where $\mathcal{F}_{\gamma_{1}}$ (blue data points) measures the average surface mass density of the filament, and $\mathcal{F}_{\gamma_{2}}$ is expected to be consistent with zero and hence serves as a null test.
    The lower right small panel shows the measurements from the three-surveys combined using inverse variance weighting. We note that this additional panel is purely for illustration, however, as our joint survey-constraints on the filament model are derived from a combination of the surveys on the likelihood-level.
    \underline{Right:} The estimated parameters in the filament density model Eq.~\ref{eqn:fitting} from the stacked signal for all surveys individually and their combination.
    \underline{Upper}: Results from $3 \sim 5\hmpc$ Physical Pairs, our filament candidates;
    \underline{Lower}: Results from $3 \sim 5\hmpc$ Non-physical Pairs, our control sample.}
    \label{fig:results}
\end{figure*}

Fig.~\ref{fig:results} presents our filament shear measurements and constraints on the two parameters of the filament model (Eq.~\ref{eqn:fitting}) for physical pairs, our filament candidates (upper panels), and non-physical pairs, our control sample (lower panels).
The left panel shows the nulled $\mathcal{F}(r)$ shear signal as a function of the distance from the centre of the filament measured in units of $\surfmass$. 
The result is presented for each lensing survey.  We also show the measurements from the three-surveys combined using inverse variance weighting.
The blue data points are a measurement of $\mathcal{F}_{\gamma_{1}}$, whereas the orange data points show the null-test $\mathcal{F}_{\gamma_{2}}$. 
The shaded region corresponds to the statistical noise from our fiducial measurements, and the capped errorbars correspond to the systematic uncertainty captured by the photometric redshift bias nuisance parameter $\delta z$ (see Sect.~\ref{sec:nz_marginalise}).
{The right panel shows the 68\% and 95\% confidence region of parameters $\mathcal{F}_c$ and $r_c$ in the \citet{mead/king/mccarthy:2010} filament model.
These estimated parameters can be compared to the best-fit parameter from the noise-free SLICS analysis in Sect.~\ref{sec:simulations} which is also represented by the cross in the right panel.
We note that the noise-free SLICS best-fit is consistent with 
{
all three surveys.
We also present joint constraints from the combined signal using a block covariance 
\begin{align}
{\rm Cov_{all}} = 
\begin{pmatrix}
  {\rm Cov_{CF}}          &0                     &0                     \\
  0                       &{\rm Cov_{RC}}        &0                     \\
  0                       &0                     &{\rm Cov_{KV}}        \\
\end{pmatrix}
\label{eqn:cov}
\end{align}
and extended data vector $\mathcal{F}_{\rm ext} = (\mathcal{F}_{\rm CF}, \mathcal{F}_{\rm RC}, \mathcal{F}_{\rm KV})^{\rm T}$,
providing an estimate of the average filament profile from all the filament candidates across the three surveys.
Eq.~\ref{eqn:cov} assumes that the surveys are uncorrelated, which is a good approximation to make given the lack of overlap between the different surveys.
}

To quantify the significance of our measurements, we use the likelihood ratio test between the null hypothesis $H_0$ and the filament model $H_1$, where the likelihood ratio $\rm LR$ is
\begin{align}
    {\rm LR}(\bm{\mathcal{F}}(r)) = 
    \frac{\sup\limits_{\theta \in B_1}\mathcal{L}(\theta|\bm{\mathcal{F}}(r))}
         {\sup\limits_{\theta \in B_0}\mathcal{L}(\theta|\bm{\mathcal{F}}(r))},
\end{align}
where $B_0$ and $B_1$ are the parameter space in each hypothesis, i.e., $B_0$ has no free parameters and $B_1 = \{\mathcal{F}_c, r_c\}$.
By Wilks' theorem \citep{wilks:1938,williams:2001}, the deviance defined as ${\rm Dev} = 2 \ln \rm LR$ has an asymptotic chi-squared distribution with $\dim(B_1) - \dim(B_0) = 2$ degrees of freedom when $H_0$ is true. Estimating the maximum likelihoods from $\chi^2_{\rm null}$ and $\chi^2_{\rm model, min}$ using Eq.~\ref{likelihood} and computing the deviance, we report the significance level for each individual survey as well as the combined analysis in Table~\ref{tab:individual_chisquares}.
{
The reported $\chi^2_{\rm min, model}$ suggests our model is a reasonable fit to the data in all cases even for KV450, where $p(\chi^2 < 7.40 | \nu=13) = 0.12$.
We find the best-fit model parameters for $\mathcal{F}_{\gamma_1}$ from all three surveys combined as $\mathcal{F}_c = 10.5 \pm 2.9 \, h \text{M}_\odot {\rm pc}^{-2}$ and $r_c = 0.4^{+0.2}_{-0.1} \hmpc$. We note that, the majority of the detection derives from CFHTLenS alone with a $3.1\,\sigma$ detection. Combining all three surveys we measure a $3.4\,\sigma$ detection of the filament weak lensing signal.
CFHTLenS is the most constraining survey as it combines both survey depth with significant SDSS overlap. 
KV450, with roughly half the source density of CFHTLenS, and RCSLenS, with roughly 20\% of the source density, will only start to add significant constraining power with the inclusion of additional overlapping SDSS area.
}

Our control sample of `Non-physical pairs' (NP) are selected to be pairs of lens galaxies with projected separations $3-5\hmpc$, but distant in redshift space ($0.033 < \Delta z < 0.04$). 
These non-physical pairs will not be connected by a filament, hence providing an important validation of our nulling approach to isolate the filament signal. We find that the measured signal for $\mathcal{F}_{\gamma_{1,2}}$ is consistent with zero for all surveys and the combined survey as shown in Table~\ref{tab:individual_chisquares} and the lower panel of Fig.~\ref{fig:results}.

{
For consistency with other analyses in the literature we also analyse 27,880 filament pairs in LOWZ that have a physical separation of $6 - 10\, \hmpc$. 
In contrast to other studies, we do not detect a significant signal for these larger-separation filaments in any of the surveys individually. In combination we find a weak signal at $1.6\, \sigma$ significance, with $\mathcal{F}_c = 1.3 \pm 0.6$.  The half-maximum radius of the density profile, $r_c$, is however unconstrained.
}
 \section{Conclusions and Discussion}
\label{discuss}
In this paper, we have presented a 
{
$3.4\,\sigma$ 
}
detection of filamentary structure connecting luminous red galaxies separated by $3-5 \hmpc$. Through a series of null tests we have verified the robustness of this result.
Our work extends the analysis presented in C16 by improving the methodology to null the weak lensing signal from the LRGs in order to isolate the weak lensing distortions induced by the filamentary structure alone. 
We note that this nulling method cannot distinguish between a pair of spherical haloes joined by a cylinder of matter, or two elliptical haloes which extend towards each other. \citet{higuchi/oguri/shirasaki:2014} shows that there is no hard line between a filament and its corresponding halo, with haloes typically extending along the filament.
As we find a strong nulled signal on scales much larger than the typical ($\sim 1\hmpc$) virial radius of the LRGs \citep{parejko/etal:2013}, we would argue that it is unlikely to originate from two perfectly aligned haloes \citep[see][]{xia/etal:2017}.
But nevertheless we prefer to refer to our detection as that of filamentary structure, rather than that of a filament per se.

Previous studies have focused on LRGs separated by $6-10\hmpc$. 
{
Using all three lensing surveys (KV450, CFHTLenS and RCSLenS), we do not detect a significant signal for these larger-separation filaments in either the surveys individually, or in combination. 
This is in contrast to the significant $5\,\sigma$ weak lensing detection of $6-10\hmpc$ separation filaments reported by \citet{epps/hudson:2017} using the same CFHTLenS dataset that has been analysed in this study.
We report that we are unable to reproduce their result, even when adopting the same methodology.
In comparison to C16, we find no comparable detection to their reported $\sim 4\,\sigma$ detection. As we have shown that sampling variance makes a significant contribution to the overall error budget, the factor of 5 increase in the number of filaments studied by C16, in contrast to this analysis, is key to their detection, even though the C16 lensing source galaxy density is significantly shallower than the source densities of KV450, CFHTLenS and RCSLenS.
Neither \citet{epps/hudson:2017} nor C16 constrain the parameters of the \citet{mead/king/mccarthy:2010} filament model, but in the case of C16 we can compare the amplitudes of the measured signals. 
When adopting the C16 nulling approach, we find the two shear measurements to be fully consistent.
The mean amplitude of our measurement is about four times larger than the amplitude reported in \citet{kondo/etal:2019} (see discussion in Appendix~\ref{app:nulling}). Given our error budget, however, the results are consistent.
}

\citet{deGraaff/etal:prep} calculate the average density $\bar{\kappa}$ between galaxy pairs separated by $6-14\hmpc$ using a CMB lensing convergence map, finding $\rho_0 \approx 5.5 \pm 2.9 \,\bar\rho(z)$.
This estimate assumes that the matter density follows a cylindrical filament model, with density
\begin{align}
    \rho(\ell, r_\perp) = \rho_0 \exp\left( -\frac{r_\perp^2}{2\sigma^2} \right) \exp\left( -\frac{\ell^2}{2\sigma^2} \right),
    \label{eqn:cylin_density}
\end{align}
where $\ell$ defines the size of the filament in the line-of-sight direction, $r_\perp$ defines the distance perpendicular to the filament axis on the projected sky and $\sigma$ is the intrinsic width.
Integrating this density model over the line-of-sight, we can relate this to the surface mass density at the centre of the filament $\mathcal{F}_c$ as 
\begin{align}
    \mathcal{F}_c = \int \rho(\ell, 0)\, {\rm d} \ell = \sqrt{2 \pi} \sigma \rho_0.
\end{align}
For the value $\sigma = 1.5 \hmpc$ adopted by \citet{deGraaff/etal:prep}, and our best-fit amplitude parameter $\mathcal{F}_c$, we find $\rho_0^{3-5} = (15.1 \pm 4.1)\, \bar\rho(z)$ for the $3-5 \hmpc$ filament sample at the average filament redshift $z = 0.299$. 
{
For the $6-10 \hmpc$ filament sample we find $\rho_0^{6-10} = (1.9 \pm 0.9)\, \bar\rho(z)$, which is consistent with the \citet{deGraaff/etal:prep} result. 
}
Adapting the \citet{mead/king/mccarthy:2010} filament model, we can also integrate the model over the perpendicular distance and calculate the total mass enclosed between the two LRGs as
\begin{align}
    M_{\rm fil}(r_c, \mathcal{F}_c, R_{\rm sep}) = R_{\rm sep} \times 2 \int_0^\infty \mathcal{F}(r) \,{\rm d}r = \pi\,r_c\,R_{\rm sep}\,\mathcal{F}_c.
\end{align}
Taking the best-fit parameters $\mathcal{F}_c$ and $r_c$ for $3-5\hmpc$ measurement, we find {$M_{\rm fil} =  4.9\pm2.0 \times 10^{13} \hsolar$}. It is worth noting that, this estimate is based on the approximation that the deflection potential vanishes along the filament major axis. A more detailed analysis would attempt to obtain the excess mass map under the framework of galaxy-galaxy-galaxy lensing \citep{simon/etal:2008,simon/etal:2019} but with a much larger separation of galaxy pairs.

Looking forward to upcoming deep weak lensing surveys such as the European Space Agency’s {\it Euclid} mission\footnote{\url{http://www.euclid-ec.org}} and the Large Synoptic Survey
Telescope (LSST\footnote{\url{http://www.lsst.org}}), and deep spectroscopic surveys such as the Dark Energy Spectroscopic Instrument (DESI\footnote{\url{http://desi.lbl.gov/}}), the methodology that we have presented in this paper could be used to probe filamentary structure as a function of LRG mass and redshift. The combination of overlapping weak lensing surveys and spectroscopic surveys will provide the optimal datasets with which to fully explore the cosmic web.
 
\section*{Acknowledgements}

We acknowledge support from the European Research Council under grant numbers 647112 (QX, CH, AA, MA, JHD, TT) and 770935 (HH, AW).
CH also acknowledges the support from the Max Planck Society and the Alexander von Humboldt Foundation in the framework of the Max Planck-Humboldt Research Award endowed by the Federal Ministry of Education and Research. 
YC acknowledges the support of the Royal Society through the award of a University Research Fellowship and an Enhancement Award.
We acknowledge support from the European Commission under a Marie-Sk{l}odwoska-Curie European Fellowship under project numbers 656869 (JHD) and 797794 (TT).
HH is supported by a Heisenberg grant of the Deutsche Forschungsgemeinschaft (Hi 1495/5-1).
AK acknowledges support from Vici grant 639.043.512, financed by the Netherlands Organisation for Scientific Research (NWO).
PS acknowledges support from the Deutsche Forschungsgemeinschaft in the framework of the TR33 `The Dark Universe'.
\newline
Computations for the $N$-body simulations were performed in part on the Orcinus supercomputer at the WestGrid HPC consortium (www.westgrid.ca), in part on the GPC supercomputer at the SciNet HPC Consortium. SciNet is funded by: the Canada Foundation for Innovation under the auspices of Compute Canada; the Government of Ontario; Ontario Research Fund - Research Excellence; and the University of Toronto.
\newline
Funding for SDSS-III has been provided by the Alfred P. Sloan Foundation, the Participating Institutions, the National Science Foundation, and the U.S. Department of Energy Office of Science. The SDSS-III web site is \url{http://www.sdss3.org/}. We thank SDSS-III for making their data products so easily accessible. 
\newline
SDSS-III is managed by the Astrophysical Research Consortium for the Participating Institutions of the SDSS-III Collaboration including the University of Arizona, the Brazilian Participation Group, Brookhaven National Laboratory, Carnegie Mellon University, University of Florida, the French Participation Group, the German Participation Group, Harvard University, the Instituto de Astrofisica de Canarias, the Michigan State/Notre Dame/JINA Participation Group, Johns Hopkins University, Lawrence Berkeley National Laboratory, Max Planck Institute for Astrophysics, Max Planck Institute for Extraterrestrial Physics, New Mexico State University, New York University, Ohio State University, Pennsylvania State University, University of Portsmouth, Princeton University, the Spanish Participation Group, University of Tokyo, University of Utah, Vanderbilt University, University of Virginia, University of Washington, and Yale University.
\newline
{\it Author contributions:} All authors contributed to the development and writing of this paper. The authorship list is given in two groups: the lead authors (QX, NR, CH), followed by an alphabetical group who contributed to either the scientific analysis or the data products.

\bibpunct{(}{)}{;}{a}{}{,}
\bibliographystyle{aa}

\begin{thebibliography}{53}
\expandafter\ifx\csname natexlab\endcsname\relax\def\natexlab#1{#1}\fi

\bibitem[{{Alam} {et~al.}(2015){Alam}, {Albareti}, {Allende Prieto}, {Anders},
  {Anderson}, {Anderton}, {Andrews}, {Armengaud}, {Aubourg}, {Bailey}, \&
  et~al.}]{alam/etal:2015}
{Alam}, S., {Albareti}, F.~D., {Allende Prieto}, C., {et~al.} 2015, \apjs, 219,
  12

\bibitem[{{Amon} {et~al.}(2018){Amon}, {Heymans}, {Klaes}, {Erben}, {Blake},
  {Hildebrandt}, {Hoekstra}, {Kuijken}, {Miller}, {Morrison}, {Choi}, {de
  Jong}, {Glazebrook}, {Irisarri}, {Joachimi}, {Joudaki}, {Kannawadi},
  {Lidman}, {Napolitano}, {Parkinson}, {Schneider}, {van Uitert}, {Viola}, \&
  {Wolf}}]{amon/etal:2018}
{Amon}, A., {Heymans}, C., {Klaes}, D., {et~al.} 2018, \mnras, 477, 4285

\bibitem[{{Arag{\'o}n-Calvo} {et~al.}(2010){Arag{\'o}n-Calvo}, {van de
  Weygaert}, \& {Jones}}]{aragoncalvo/van/jones:2010}
{Arag{\'o}n-Calvo}, M.~A., {van de Weygaert}, R., \& {Jones}, B.~J.~T. 2010,
  \mnras, 408, 2163

\bibitem[{{Bartelmann} \& {Schneider}(2001)}]{bartelmann/schneider:2001}
{Bartelmann}, M. \& {Schneider}, P. 2001, Physics Reports, 340, 291

\bibitem[{{Ben{\'{\i}}tez}(2000)}]{benitez:2000}
{Ben{\'{\i}}tez}, N. 2000, \apj, 536, 571

\bibitem[{{Bond} {et~al.}(1996){Bond}, {Kofman}, \&
  {Pogosyan}}]{bond/kofman/pogosyan:1996}
{Bond}, J.~R., {Kofman}, L., \& {Pogosyan}, D. 1996, \nat, 380, 603

\bibitem[{{Bregman}(2007)}]{bregman:2007}
{Bregman}, J.~N. 2007, \araa, 45, 221

\bibitem[{{Briel} \& {Henry}(1995)}]{briel/henry:1995}
{Briel}, U.~G. \& {Henry}, J.~P. 1995, Astronomy and Astrophysics, 302, L9

\bibitem[{{Cheng} \& {Gupta}(1989)}]{cheng/gupta:1989}
{Cheng}, H. \& {Gupta}, K.~C. 1989, ASME Transactions Series E Journal of
  Applied Mechanics, 56, 139

\bibitem[{{Choi} {et~al.}(2016){Choi}, {Heymans}, {Blake}, {Hildebrandt},
  {Duncan}, {Erben}, {Nakajima}, {Van Waerbeke}, \& {Viola}}]{choi/etal:2016}
{Choi}, A., {Heymans}, C., {Blake}, C., {et~al.} 2016, \mnras, 463, 3737

\bibitem[{{Clampitt} {et~al.}(2016){Clampitt}, {Miyatake}, {Jain}, \&
  {Takada}}]{clampitt/etal:2016}
{Clampitt}, J., {Miyatake}, H., {Jain}, B., \& {Takada}, M. 2016, \mnras, 457,
  2391

\bibitem[{{Colberg} {et~al.}(2005){Colberg}, {Krughoff}, \&
  {Connolly}}]{colberg/krughoff/connolly:2005}
{Colberg}, J.~M., {Krughoff}, K.~S., \& {Connolly}, A.~J. 2005, \mnras, 359,
  272

\bibitem[{{Colless} {et~al.}(2001){Colless}, {Dalton}, {Maddox}, {Sutherland},
  {Norberg}, {Cole}, {Bland-Hawthorn}, {Bridges}, {Cannon}, {Collins}, {Couch},
  {Cross}, {Deeley}, {De Propris}, {Driver}, {Efstathiou}, {Ellis}, {Frenk},
  {Glazebrook}, {Jackson}, {Lahav}, {Lewis}, {Lumsden}, {Madgwick}, {Peacock},
  {Peterson}, {Price}, {Seaborne}, \& {Taylor}}]{colless/etal:2001}
{Colless}, M., {Dalton}, G., {Maddox}, S., {et~al.} 2001, \mnras, 328, 1039

\bibitem[{{de Graaff} {et~al.}(2019){de Graaff}, {Cai}, {Heymans}, \&
  {Peacock}}]{deGraaff/etal:prep}
{de Graaff}, A., {Cai}, Y.-C., {Heymans}, C., \& {Peacock}, J.~A. 2019,
  Astronomy {\&} Astrophysics, 624, A48

\bibitem[{{Dietrich} {et~al.}(2012){Dietrich}, {Werner}, {Clowe}, {Finoguenov},
  {Kitching}, {Miller}, \& {Simionescu}}]{dietrich/etal:2012}
{Dietrich}, J.~P., {Werner}, N., {Clowe}, D., {et~al.} 2012, \nat, 487, 202

\bibitem[{{Dolag} {et~al.}(2006){Dolag}, {Meneghetti}, {Moscardini}, {Rasia},
  \& {Bonaldi}}]{dolag/etal:2006}
{Dolag}, K., {Meneghetti}, M., {Moscardini}, L., {Rasia}, E., \& {Bonaldi}, A.
  2006, \mnras, 370, 656

\bibitem[{{Dvornik} {et~al.}(2018){Dvornik}, {Hoekstra}, {Kuijken},
  {Schneider}, {Amon}, {Nakajima}, {Viola}, {Choi}, {Erben}, {Farrow},
  {Heymans}, {Hildebrandt}, {Sif{\'o}n}, \& {Wang}}]{dvornik/etal:2018}
{Dvornik}, A., {Hoekstra}, H., {Kuijken}, K., {et~al.} 2018, \mnras, 479, 1240

\bibitem[{{Eardley} {et~al.}(2015){Eardley}, {Peacock}, {McNaught-Roberts},
  {Heymans}, {Norberg}, {Alpaslan}, {Baldry}, {Bland-Hawthorn}, {Brough},
  {Cluver}, {Driver}, {Farrow}, {Liske}, {Loveday}, \&
  {Robotham}}]{eardley/etal:2015}
{Eardley}, E., {Peacock}, J.~A., {McNaught-Roberts}, T., {et~al.} 2015, \mnras,
  448, 3665

\bibitem[{{Ebeling} {et~al.}(2004){Ebeling}, {Barrett}, \&
  {Donovan}}]{ebeling/barrett/donovan:2004}
{Ebeling}, H., {Barrett}, E., \& {Donovan}, D. 2004, \apjl, 609, L49

\bibitem[{{Epps} \& {Hudson}(2017)}]{epps/hudson:2017}
{Epps}, S.~D. \& {Hudson}, M.~J. 2017, \mnras, 468, 2605

\bibitem[{{Harnois-D{\'e}raps} {et~al.}(2018){Harnois-D{\'e}raps}, {Amon},
  {Choi}, {Demchenko}, {Heymans}, {Kannawadi}, {Nakajima}, {Sirks}, {van
  Waerbeke}, {Cai}, {Giblin}, {Hildebrandt}, {Hoekstra}, {Miller}, \&
  {Tr{\"o}ster}}]{harnoisderaps/etal:2018}
{Harnois-D{\'e}raps}, J., {Amon}, A., {Choi}, A., {et~al.} 2018, \mnras, 481,
  1337

\bibitem[{{Harnois-D{\'e}raps} \& {van
  Waerbeke}(2015)}]{harnoisderaps/van:2015}
{Harnois-D{\'e}raps}, J. \& {van Waerbeke}, L. 2015, \mnras, 450, 2857

\bibitem[{{Heymans} {et~al.}(2006){Heymans}, {Van Waerbeke}, {Bacon}, {Berge},
  {Bernstein}, {Bertin}, {Bridle}, {Brown}, {Clowe}, {Dahle}, {Erben}, {Gray},
  {Hetterscheidt}, {Hoekstra}, {Hudelot}, {Jarvis}, {Kuijken}, {Margoniner},
  {Massey}, {Mellier}, {Nakajima}, {Refregier}, {Rhodes}, {Schrabback}, \&
  {Wittman}}]{heymans/etal:2006}
{Heymans}, C., {Van Waerbeke}, L., {Bacon}, D., {et~al.} 2006, \mnras, 368,
  1323

\bibitem[{{Heymans} {et~al.}(2012){Heymans}, {Van Waerbeke}, {Miller}, {Erben},
  {Hildebrandt}, {Hoekstra}, {Kitching}, {Mellier}, {Simon}, {Bonnett},
  {Coupon}, {Fu}, {Harnois D{\'e}raps}, {Hudson}, {Kilbinger}, {Kuijken},
  {Rowe}, {Schrabback}, {Semboloni}, {van Uitert}, {Vafaei}, \&
  {Velander}}]{heymans/etal:2012}
{Heymans}, C., {Van Waerbeke}, L., {Miller}, L., {et~al.} 2012, \mnras, 427,
  146

\bibitem[{{Higuchi} {et~al.}(2014){Higuchi}, {Oguri}, \&
  {Shirasaki}}]{higuchi/oguri/shirasaki:2014}
{Higuchi}, Y., {Oguri}, M., \& {Shirasaki}, M. 2014, \mnras, 441, 745

\bibitem[{{Hildebrandt} {et~al.}(2016){Hildebrandt}, {Choi}, {Heymans},
  {Blake}, {Erben}, {Miller}, {Nakajima}, {van Waerbeke}, {Viola},
  {Buddendiek}, {Harnois-D{\'e}raps}, {Hojjati}, {Joachimi}, {Joudaki},
  {Kitching}, {Wolf}, {Gwyn}, {Johnson}, {Kuijken}, {Sheikhbahaee}, {Tudorica},
  \& {Yee}}]{hildebrandt/etal:2016}
{Hildebrandt}, H., {Choi}, A., {Heymans}, C., {et~al.} 2016, \mnras, 463, 635

\bibitem[{{Hildebrandt} {et~al.}(2012){Hildebrandt}, {Erben}, {Kuijken}, {van
  Waerbeke}, {Heymans}, {Coupon}, {Benjamin}, {Bonnett}, {Fu}, {Hoekstra},
  {Kitching}, {Mellier}, {Miller}, {Velander}, {Hudson}, {Rowe}, {Schrabback},
  {Semboloni}, \& {Ben{\'{\i}}tez}}]{hildebrandt/etal:2012}
{Hildebrandt}, H., {Erben}, T., {Kuijken}, K., {et~al.} 2012, \mnras, 421, 2355

\bibitem[{{Hildebrandt} {et~al.}(2018){Hildebrandt}, {K{\"o}hlinger}, {van den
  Busch}, {Joachimi}, {Heymans}, {Kannawadi}, {Wright}, {Asgari}, {Blake},
  {Hoekstra}, {Joudaki}, {Kuijken}, {Miller}, {Morrison}, {Tr{\"o}ster},
  {Amon}, {Archidiacono}, {Brieden}, {Choi}, {de Jong}, {Erben}, {Giblin},
  {Mead}, {Peacock}, {Radovich}, {Schneider}, {Sif{\'o}n}, \&
  {Tewes}}]{hildebrandt/etal:2018}
{Hildebrandt}, H., {K{\"o}hlinger}, F., {van den Busch}, J.~L., {et~al.} 2018,
  arXiv e-prints [\eprint[arXiv]{1812.06076}]

\bibitem[{{Hildebrandt} {et~al.}(2017){Hildebrandt}, {Viola}, {Heymans},
  {Joudaki}, {Kuijken}, {Blake}, {Erben}, {Joachimi}, {Klaes}, {Miller},
  {Morrison}, {Nakajima}, {Verdoes Kleijn}, {Amon}, {Choi}, {Covone}, {de
  Jong}, {Dvornik}, {Fenech Conti}, {Grado}, {Harnois-D{\'e}raps}, {Herbonnet},
  {Hoekstra}, {K{\"o}hlinger}, {McFarland}, {Mead}, {Merten}, {Napolitano},
  {Peacock}, {Radovich}, {Schneider}, {Simon}, {Valentijn}, {van den Busch},
  {van Uitert}, \& {Van Waerbeke}}]{hildebrandt/etal:2017}
{Hildebrandt}, H., {Viola}, M., {Heymans}, C., {et~al.} 2017, \mnras, 465, 1454

\bibitem[{{Kondo} {et~al.}(2019){Kondo}, {Miyatake}, {Shirasaki}, {Sugiyama},
  \& {Nishizawa}}]{kondo/etal:2019}
{Kondo}, H., {Miyatake}, H., {Shirasaki}, M., {Sugiyama}, N., \& {Nishizawa},
  A.~J. 2019, arXiv e-prints [\eprint[arXiv]{1905.08991}]

\bibitem[{{Kull} \& {B{\"o}hringer}(1999)}]{kull/bohringer:1999}
{Kull}, A. \& {B{\"o}hringer}, H. 1999, \aap, 341, 23

\bibitem[{{Macci{\`o}} {et~al.}(2007){Macci{\`o}}, {Dutton}, {van den Bosch},
  {Moore}, {Potter}, \& {Stadel}}]{maccio/etal:2007}
{Macci{\`o}}, A.~V., {Dutton}, A.~A., {van den Bosch}, F.~C., {et~al.} 2007,
  \mnras, 378, 55

\bibitem[{{Mandelbaum} {et~al.}(2005){Mandelbaum}, {Hirata}, {Seljak}, {Guzik},
  {Padmanabhan}, {Blake}, {Blanton}, {Lupton}, \&
  {Brinkmann}}]{mandelbaum/etal:2005}
{Mandelbaum}, R., {Hirata}, C.~M., {Seljak}, U., {et~al.} 2005, \mnras, 361,
  1287

\bibitem[{{Mead} {et~al.}(2010){Mead}, {King}, \&
  {McCarthy}}]{mead/king/mccarthy:2010}
{Mead}, J.~M.~G., {King}, L.~J., \& {McCarthy}, I.~G. 2010, \mnras, 401, 2257

\bibitem[{{Miller} {et~al.}(2013){Miller}, {Heymans}, {Kitching}, {van
  Waerbeke}, {Erben}, {Hildebrandt}, {Hoekstra}, {Mellier}, {Rowe}, {Coupon},
  {Dietrich}, {Fu}, {Harnois-D{\'e}raps}, {Hudson}, {Kilbinger}, {Kuijken},
  {Schrabback}, {Semboloni}, {Vafaei}, \& {Velander}}]{miller/etal:2013}
{Miller}, L., {Heymans}, C., {Kitching}, T.~D., {et~al.} 2013, \mnras, 429,
  2858

\bibitem[{{Parejko} {et~al.}(2013){Parejko}, {Sunayama}, {Padmanabhan}, {Wake},
  {Berlind}, {Bizyaev}, {Blanton}, {Bolton}, {van den Bosch}, {Brinkmann},
  {Brownstein}, {da Costa}, {Eisenstein}, {Guo}, {Kazin}, {Maia},
  {Malanushenko}, {Maraston}, {McBride}, {Nichol}, {Oravetz}, {Pan},
  {Percival}, {Prada}, {Ross}, {Ross}, {Schlegel}, {Schneider}, {Simmons},
  {Skibba}, {Tinker}, {Tojeiro}, {Weaver}, {Wetzel}, {White}, {Weinberg},
  {Thomas}, {Zehavi}, \& {Zheng}}]{parejko/etal:2013}
{Parejko}, J.~K., {Sunayama}, T., {Padmanabhan}, N., {et~al.} 2013, \mnras,
  429, 98

\bibitem[{{Pimbblet} \& {Drinkwater}(2004)}]{pimbblet/drinkwater:2004}
{Pimbblet}, K.~A. \& {Drinkwater}, M.~J. 2004, \mnras, 347, 137

\bibitem[{{Ross} {et~al.}(2012){Ross}, {Percival}, {S{\'a}nchez}, {Samushia},
  {Ho}, {Kazin}, {Manera}, {Reid}, {White}, {Tojeiro}, {McBride}, {Xu}, {Wake},
  {Strauss}, {Montesano}, {Swanson}, {Bailey}, {Bolton}, {Dorta}, {Eisenstein},
  {Guo}, {Hamilton}, {Nichol}, {Padmanabhan}, {Prada}, {Schlegel},
  {Maga{\~n}a}, {Zehavi}, {Blanton}, {Bizyaev}, {Brewington}, {Cuesta},
  {Malanushenko}, {Malanushenko}, {Oravetz}, {Parejko}, {Pan}, {Schneider},
  {Shelden}, {Simmons}, {Snedden}, \& {Zhao}}]{ross/etal:2012}
{Ross}, A.~J., {Percival}, W.~J., {S{\'a}nchez}, A.~G., {et~al.} 2012, \mnras,
  424, 564

\bibitem[{{Schneider} \& {Watts}(2005)}]{schneider/watts:2005}
{Schneider}, P. \& {Watts}, P. 2005, \aap, 432, 783

\bibitem[{{Simon} {et~al.}(2019){Simon}, {Saghiha}, {Hilbert}, {Schneider},
  {Boever}, \& {Wright}}]{simon/etal:2019}
{Simon}, P., {Saghiha}, H., {Hilbert}, S., {et~al.} 2019, \aap, 622, A104

\bibitem[{{Simon} {et~al.}(2008){Simon}, {Watts}, {Schneider}, {Hoekstra},
  {Gladders}, {Yee}, {Hsieh}, \& {Lin}}]{simon/etal:2008}
{Simon}, P., {Watts}, P., {Schneider}, P., {et~al.} 2008, \aap, 479, 655

\bibitem[{{Springel} {et~al.}(2005){Springel}, {White}, {Jenkins}, {Frenk},
  {Yoshida}, {Gao}, {Navarro}, {Thacker}, {Croton}, {Helly}, {Peacock}, {Cole},
  {Thomas}, {Couchman}, {Evrard}, {Colberg}, \& {Pearce}}]{springel/etal:2005}
{Springel}, V., {White}, S.~D.~M., {Jenkins}, A., {et~al.} 2005, \nat, 435, 629

\bibitem[{{Tanimura} {et~al.}(2019){Tanimura}, {Hinshaw}, {McCarthy}, {Van
  Waerbeke}, {Aghanim}, {Ma}, {Mead}, {Hojjati}, \&
  {Tr{\"o}ster}}]{tanimura/etal:prep}
{Tanimura}, H., {Hinshaw}, G., {McCarthy}, I.~G., {et~al.} 2019, \mnras, 483,
  223

\bibitem[{{Velander} {et~al.}(2011){Velander}, {Kuijken}, \&
  {Schrabback}}]{velander/kuijken/schrabback:2011}
{Velander}, M., {Kuijken}, K., \& {Schrabback}, T. 2011, \mnras, 412, 2665

\bibitem[{{Velander} {et~al.}(2014){Velander}, {van Uitert}, {Hoekstra},
  {Coupon}, {Erben}, {Heymans}, {Hildebrandt}, {Kitching}, {Mellier}, {Miller},
  {Van Waerbeke}, {Bonnett}, {Fu}, {Giodini}, {Hudson}, {Kuijken}, {Rowe},
  {Schrabback}, \& {Semboloni}}]{velander/etal:2014}
{Velander}, M., {van Uitert}, E., {Hoekstra}, H., {et~al.} 2014, \mnras, 437,
  2111

\bibitem[{{Werner} {et~al.}(2008){Werner}, {Finoguenov}, {Kaastra},
  {Simionescu}, {Dietrich}, {Vink}, \& {B{\"o}hringer}}]{werner/etal:2008}
{Werner}, N., {Finoguenov}, A., {Kaastra}, J.~S., {et~al.} 2008, \aap, 482, L29

\bibitem[{Wilks(1938)}]{wilks:1938}
Wilks, S.~S. 1938, Ann. Math. Statist., 9, 60

\bibitem[{Williams(2001)}]{williams:2001}
Williams, D. 2001, Weighing the Odds: A Course in Probability and Statistics
  (Cambridge University Press)

\bibitem[{{Wright} {et~al.}(2018){Wright}, {Hildebrandt}, {Kuijken}, {Erben},
  {Blake}, {Buddelmeijer}, {Choi}, {Cross}, {de Jong}, {Edge},
  {Gonzalez-Fernandez}, {Gonz{\'a}lez Solares}, {Grado}, {Heymans}, {Irwin},
  {Kupcu Yoldas}, {Lewis}, {Mann}, {Napolitano}, {Radovich}, {Schneider},
  {Sif{\'o}n}, {Sutherland}, {Sutorius}, \& {Verdoes
  Kleijn}}]{wright/etal:2018}
{Wright}, A.~H., {Hildebrandt}, H., {Kuijken}, K., {et~al.} 2018, arXiv
  e-prints [\eprint[arXiv]{1812.06077}]

\bibitem[{{Wright} \& {Brainerd}(2000)}]{wright/brainerd:2000}
{Wright}, C.~O. \& {Brainerd}, T.~G. 2000, \apj, 534, 34

\bibitem[{{Xia} {et~al.}(2017){Xia}, {Kang}, {Wang}, {Luo}, {Yang}, {Jing},
  {Wang}, \& {Mo}}]{xia/etal:2017}
{Xia}, Q., {Kang}, X., {Wang}, P., {et~al.} 2017, \apj, 848, 22

\bibitem[{{Zehavi} {et~al.}(2011){Zehavi}, {Zheng}, {Weinberg}, {Blanton},
  {Bahcall}, {Berlind}, {Brinkmann}, {Frieman}, {Gunn}, {Lupton}, {Nichol},
  {Percival}, {Schneider}, {Skibba}, {Strauss}, {Tegmark}, \&
  {York}}]{zehavi/etal:2011}
{Zehavi}, I., {Zheng}, Z., {Weinberg}, D.~H., {et~al.} 2011, \apj, 736, 59

\bibitem[{{Zel'dovich}(1970)}]{Zel'dovich:1970}
{Zel'dovich}, Y.~B. 1970, \aap, 5, 84

\end{thebibliography}

\appendix
\section{Remarks on the nulling technique}
\label{app:nulling}

In this Appendix we review the C16 nulling technique and develop an improved methodology to isolate the weak lensing signal from filaments. 
To explain the motivation behind nulling, we start with a single circularly symmetric halo positioned at the origin (0,0), for which the complex shear is given by
\begin{align}
    \gamma(\bm{r}) = \gamma(r,\theta) = \gamma_1 + {\rm{i}} \gamma_2 = -(\bar\kappa-\kappa) e^{2\rm{i}\theta} ,
\end{align}
where $\bar\kappa$ is the mean convergence inside $r$.
We define its counterpart $\gamma^c$ as the complex shear at the same radial position, with a clockwise rotation of $90^\circ$, such that
\begin{align}
    \gamma^c(\bm{r}) \equiv \gamma\left(r,\theta+\frac{\pi}{2}\right) = -\gamma(r,\theta).
\end{align}
The counterpart is therefore able to ``null'' the shear, as $\gamma^c(\bm{r}) + \gamma(\bm{r}) = 0$. 

\begin{figure}[h]
    \centering
    \includegraphics[width=\columnwidth]{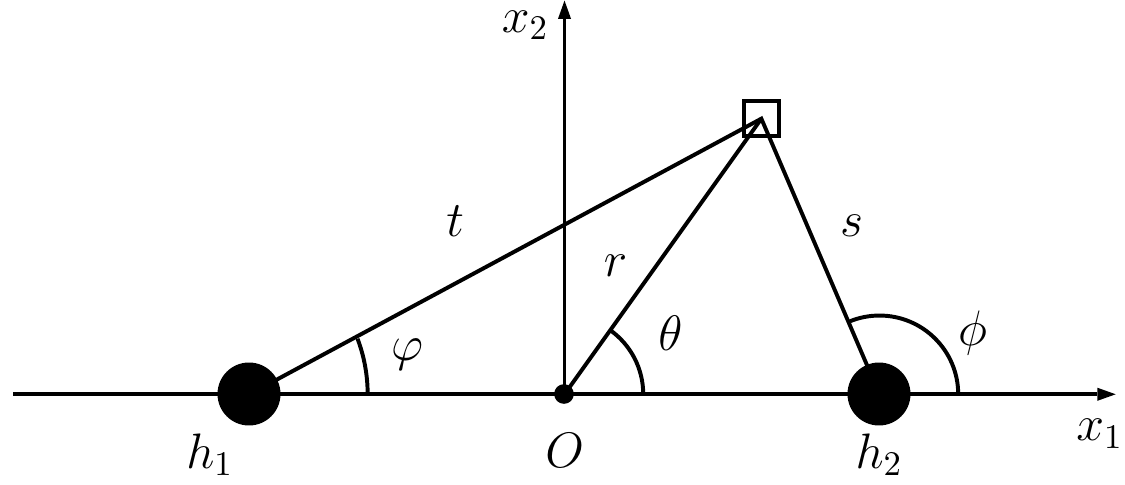}
    \caption{The bipolar configuration for two haloes centred symmetrically about origin $O$ at $h_1$ and $h_2$.}
    \label{fig:bipolar}
\end{figure}

\begin{figure}[h]
    \centering
    \includegraphics[width=1.\columnwidth]{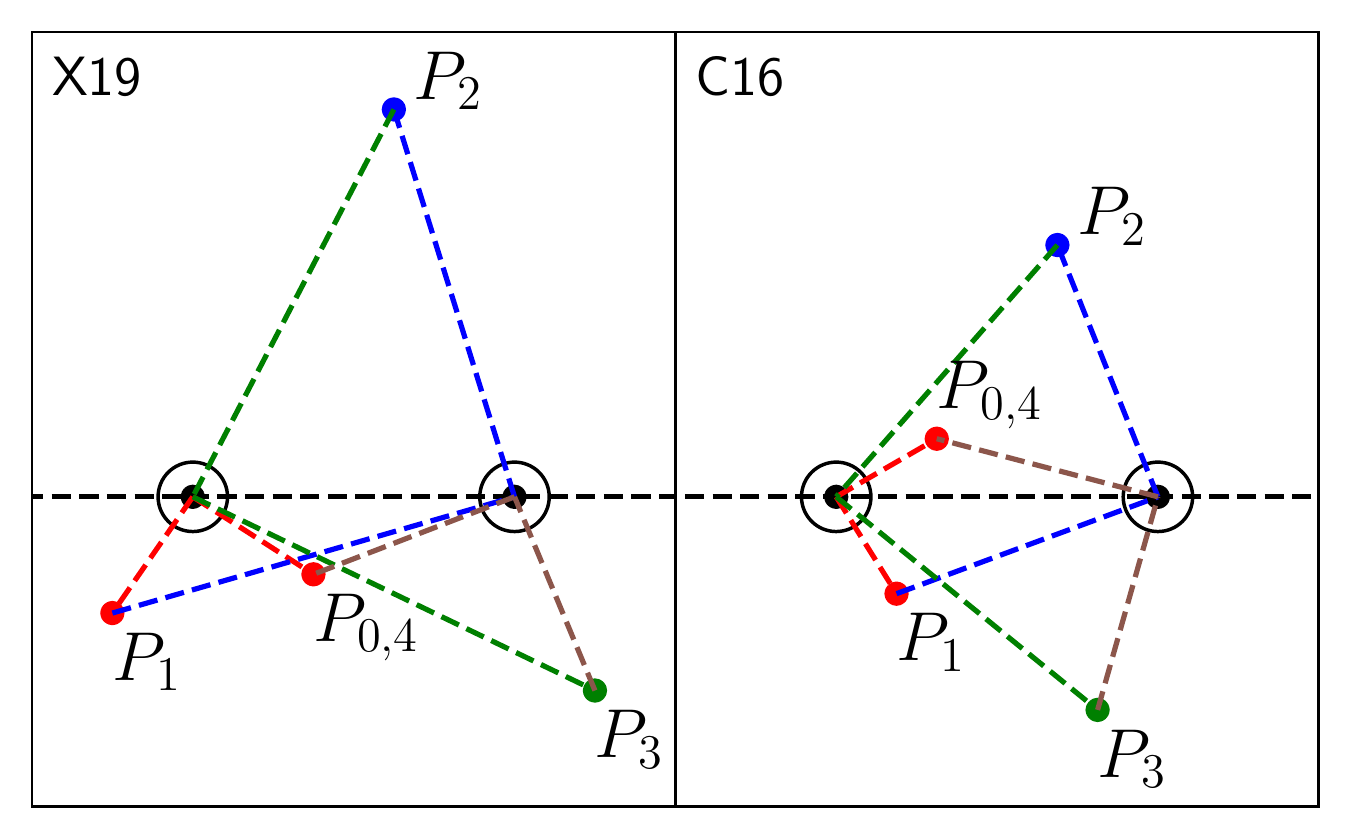}
    \caption{Illustration of the nulling technique. Every two line segments with the same colour represents an anti-clockwise rotation with respect to one halo. Right: The configuration described in C16. Left: Our adopted ``X19'' configuration which starts below the horizontal axis.}
    \label{fig:Nulling_Proof}
\end{figure}
For a two-halo system as shown in Fig.~\ref{fig:bipolar}, the shear at each position $(r, \theta)$ is composed with the shear from halo $h_1$ and the shear from halo $h_2$. 
We can write this as
\begin{align}
    \gamma({r, \theta}) &= \gamma_{h_1}(t,\varphi) + \gamma_{h_2}(s,\phi)
\end{align}
where the co-ordinates $(t, \varphi)$ are defined with halo $h_1$ at the origin, and the co-ordinates $(s, \phi)$ are defined with halo $h_2$ at the origin.
Starting from position $P_0=(r_0, \theta_0)$, shear is given by
\begin{align}
    P_0 : \gamma({r_0, \theta_0}) &= \gamma_{h_1}(t_0,\varphi_0) + \gamma_{h_2}(s_0,\phi_0).
\end{align}
A clockwise rotation around halo $h_1$ by $90^\circ$ takes us to position $P_1$ at $(r_1,\theta_1)$. The shear here is given by
\begin{align}
    P_1 : \gamma({r_1, \theta_1}) &= \gamma_{h_1}\left(t_0,\varphi_0 + \frac\pi2\right) + \gamma_{h_2}(s_1,\phi_1).
\end{align}
At position $P_1$, we see the shear contribution from halo $h_1$ is the counterpart to the shear contribution from halo $h_1$ at position $P_0$.
We next rotate around halo $h_2$ by $90^\circ$ to position $P_2 = (r_2, \theta_2)$, where the shear is given by
\begin{align}
    P_2 : \gamma({r_2, \theta_2}) &= \gamma_{h_1}(t_2,\varphi_2) + \gamma_{h_2}\left(s_1,\phi_1+\frac\pi2\right).
\end{align}
Similarly another $90^\circ$ rotation about halo $h_1$ (see Fig.~\ref{fig:Nulling_Proof}) to position $P_3$ gives 
\begin{align}
    P_3 : \gamma({r_3, \theta_3}) &= \gamma_{h_1}\left(t_2,\varphi_2 + \frac\pi2\right) + \gamma_{h_2}(s_3,\phi_3).
\end{align}
We note that, after another $90^\circ$ rotation about halo $h_2$, we come to position $P_4$ which is our starting point $P_0$. 
The sum of the shear from position $P_0, P_1, P_2$ and $P_3$ is given by 
\begin{align}
    \sum_{i=0}^3\gamma(r_i, \theta_i) &= \gamma_{h_2}(s_0,\phi_0) + \gamma_{h_2}(s_3,\phi_3) \\ 
                                      &= \gamma_{h_2}\left(s_3,\phi_3 + \frac\pi2\right) + \gamma_{h_2}(s_3,\phi_3) = 0.
\end{align}
If we now add in a filament shear $\gamma_{\rm f}$ such that at each position $\gamma = \gamma_{\rm f} + \gamma_{h_1} + \gamma_{h_2}$, then 
\begin{align}
    \sum_{i=0}^3 \gamma(r_i,\theta_i) = \gamma_{\rm f}(r_0,\theta_0) + \gamma_{\rm f}(r_1,\theta_1) + \gamma_{\rm f}(r_2,\theta_2) + \gamma_{\rm f}(r_3,\theta_3).
    \label{eqn:A9}
\end{align}
{ As we expect the filament shear profile to be symmetric about the horizontal axis, we also sum over the shear from positions $P_0', P_1', P_2'$ and $P_3'$ that are reflections of $P_0, P_1, P_2$ and $P_3$ about the horizontal axis respectively, in order to get the average shear value at any distance away from filament axis. }
We therefore define the nulling operator $\mathcal{N}$ as 
\begin{align}
    \mathcal{N}[\gamma(r_0, \theta_0)] = \frac12 \left(\sum_{i=0}^3\gamma(r_i, \theta_i) + \sum_{i=0}^3\gamma(r_i', \theta_i')\right) 
    \label{eqn:nulling}
\end{align}
It is interesting to note that the above equations also apply when two halos are of different masses given their circular symmetry.

\begin{figure}
    \centering
    \includegraphics[width=0.7\columnwidth]{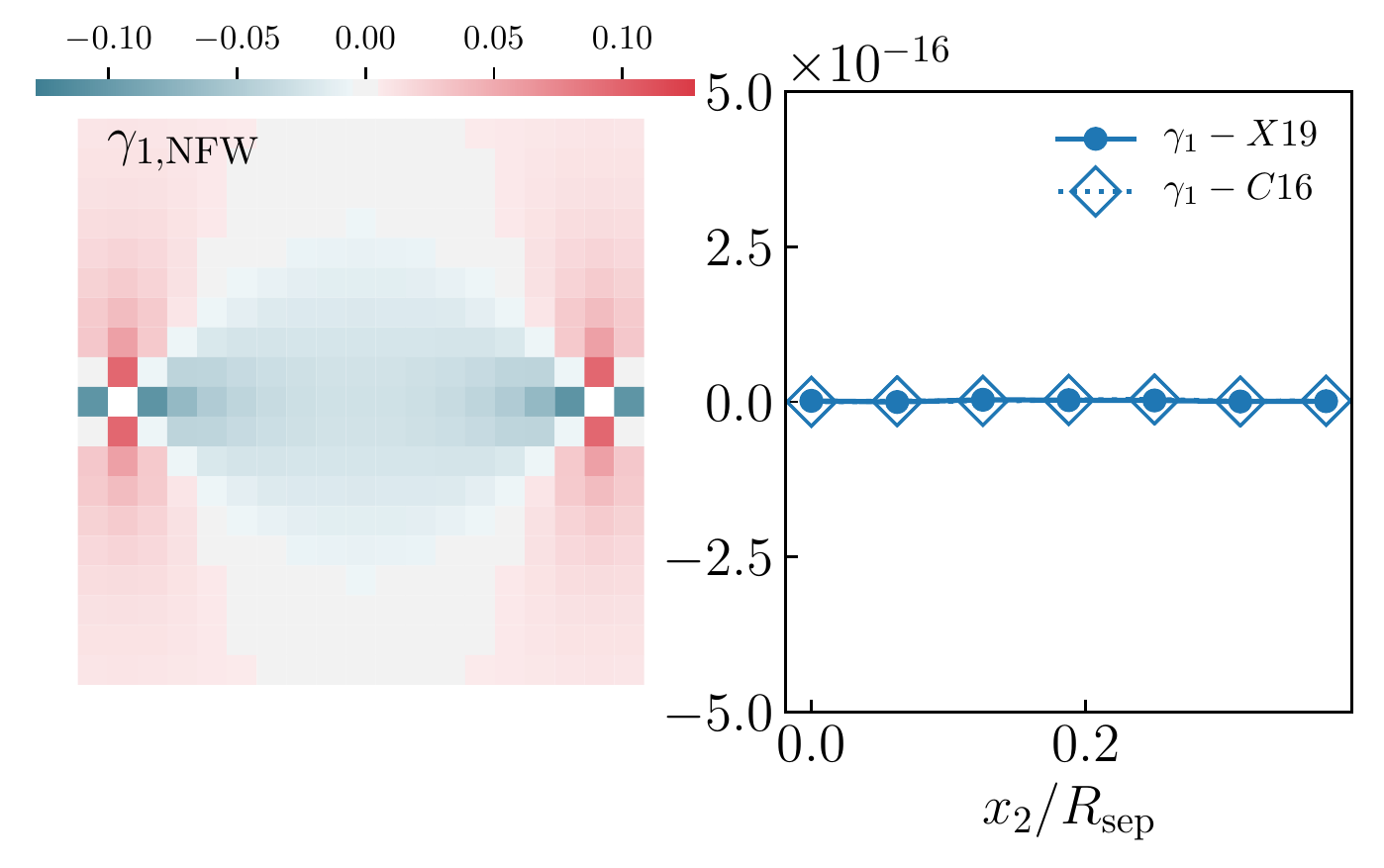}
    \includegraphics[width=0.7\columnwidth]{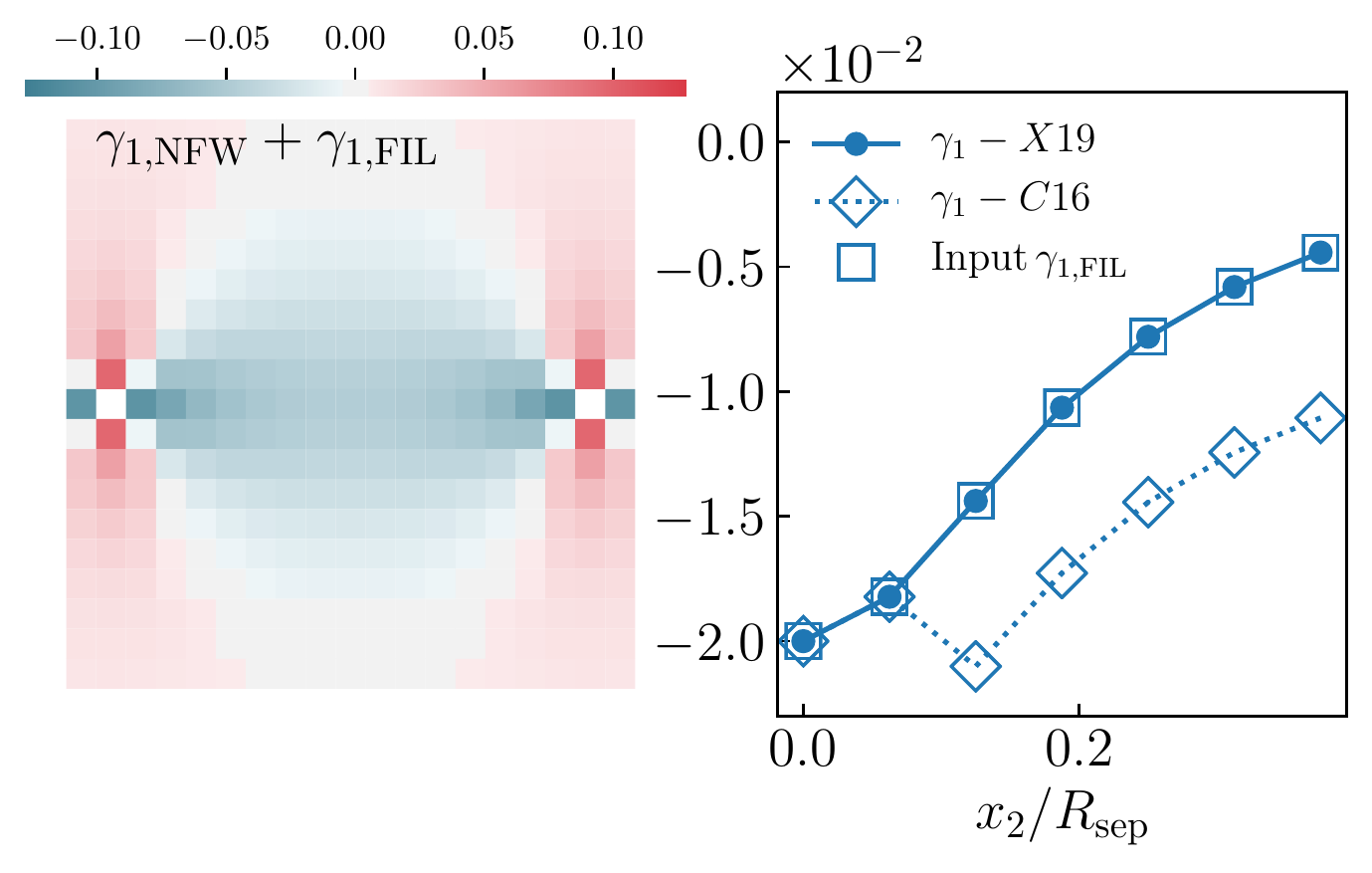}
    \includegraphics[width=0.7\columnwidth]{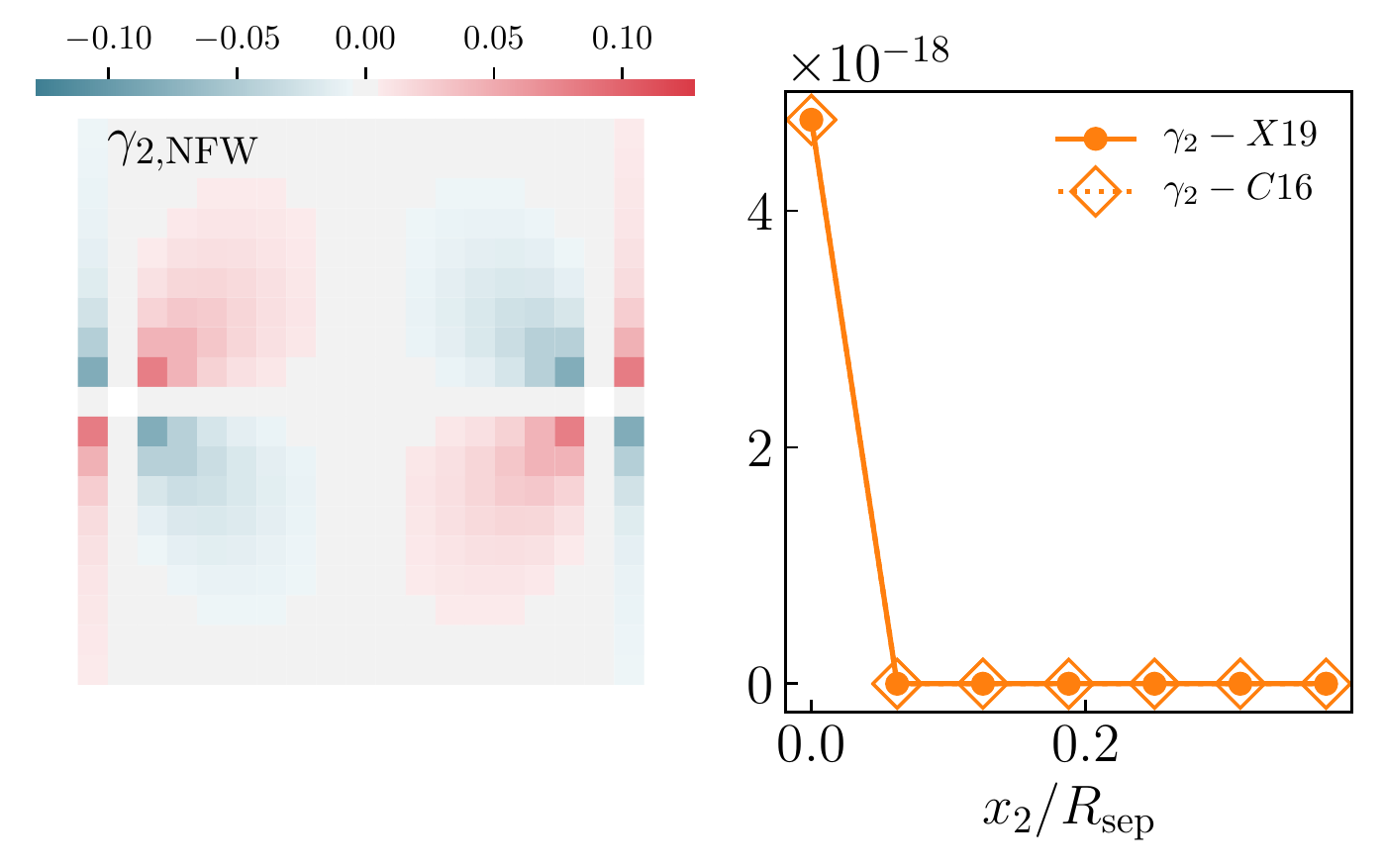}
    \caption{\underline{Left panel}: Shear maps from two NFW halos with or without fiducial filaments. The top row shows the $\gamma_1$ map generated by two NFW haloes only, while the middle row shows the $\gamma_1$ map with the addition of a fiducial filament profile. The lowest row shows the $\gamma_2$ map generated by two NFW haloes. \underline{Right panel}: Results of the nulling procedure corresponding to the shear map on the left. In the top row, we see under both the X19 configuration (solid line) and C16's (dashed line), the resulting signal is consistent with zero. When adding a fiducial filament profile, in the middle row, we see our X19 configuration correctly recovers the input value whereas the C16 configuration is biased on large scales. In the lower row, we verify that both configurations null the $\gamma_2$ signal from two NFW haloes.}
    \label{fig:NFW_on_off_FIL}
\end{figure}

In Fig.~\ref{fig:Nulling_Proof} we show two configurations for nulling. C16 chose to start $P_0$ above the horizontal axis and our work starts $P_0$ below the horizontal axis.
We investigate the difference between these two approaches by first constructing a noiseless shear map from two NFW halo profiles using Eq.~14 in \citet{wright/brainerd:2000} with $M_{\rm vir} = 10^{13} \hsolar$ and assuming a mass-concentration relation from \citet{maccio/etal:2007}. 
We assume both haloes are located at $z = 0.3$ with background sources at redshift 0.7, close to the mean value of KiDS. 
The resultant $\gamma_1$ map is shown in the top-left panel of Fig.~\ref{fig:NFW_on_off_FIL}.

For each pixel $(x_1,x_2)$ on the map, we calculate $\mathcal{N}[\gamma_{i}(x_1,x_2)]$, and show in the top-right panel in Fig.~\ref{fig:NFW_on_off_FIL}, the average of the sum along the horizontal axis, i.e., $\gamma_1^{\rm null}(r) = \sum\limits_{x=x_{\rm min}}^{x_{\rm max}} \mathcal{N} [\gamma_1(x_1, x_2)]/(x_{\rm max} - x_{\rm min} + 1)$.
{ In the lower-right panel in Fig.~\ref{fig:NFW_on_off_FIL}, we show the resulting $\mathcal{N}[\gamma_{2}(x_1,x_2)]$.}
As expected, the sum of nulling pixels are zeros (note the ${10^{-16}/10^{-18}}$ on the $y$-axis).
{In the middle panel we repeat the analysis with the inclusion of a fiducial filament modelled using the power-law profile model in Eq.~\ref{eqn:fitting}. The filament contribution to the shear $\gamma_{1}^{\rm fil}(r)$ is included in the 2D shear map as a power-law, symmetric about the $x_1$-axis (at $x_2=0$) with $k_c=0.02$ and $r_c=0.4\hmpc$. These parameters were chosen to roughly replicate previous studies \citep{dolag/etal:2006}, in order to model the true signal contrast in observations. If the nulling method is precise, we would expect the nulled filament signal measured from the map, $\gamma_{1}^{\rm null}(r)$ to recover the input shear from the filament, $\gamma_{1}^{\rm fil}(r)$, irrespective of the nulling method used. The middle-right panel shows that this is indeed the case, with our nulling method.  The nulling method in C16, however, recovers a lower expectation value for the shear induced by the filament.}
{ 
This difference results from the C16 configuration. 
As shown in Eq.~\ref{eqn:A9}, for the C16 configuration, since $P_0$ and $P_1$ are both in the filament region, the nulling operator effectively mixes scales. This has an effect in producing a significant signal on large scales which does not reflect the underlying filament density profile. 
As $P_1, P_2$ and $P_3$ in our adopted configuration lies outside the bridge between two haloes, $\gamma_{\rm f}(r_1,\theta_1) + \gamma_{\rm f}(r_2,\theta_2) + \gamma_{\rm f}(r_3,\theta_3)$ is negligible. This enables us to recover the density profile accurately.
}
{
We note that \citet{kondo/etal:2019} adopted the C16 estimator but in their Eq.~9 (the equivalent of our Eq.~\ref{eqn:nulling}), they included an additional factor of 4 in the denominator, which, in our test-case in Fig.~\ref{fig:NFW_on_off_FIL} would result in the underestimation of the filament signal by a factor of 4.
} \section{Remarks on the spherical rotation}
\label{app:sph_rot}

Here we detail the spherical rotation used in Sect. 
\ref{Method} to project all filaments onto the same reference frame.
As illustrated in Fig.~\ref{fig:sph_rot}, we rotate all galaxies about a given axis such that the filament (the shorter arc connected by solid pink diamonds) is transformed to lie horizontally on the equator (hollow pink diamonds). 
To do this, we first transfer the right ascension and declination onto a 3D vector on a unit sphere, such that their positions are $\bm{g_1}$ and $\bm{g_2}$. 
The normal vector is defined as $\bm{\hat{n}} = \bm{g_1} \times \bm{g_2}$. 
Noting the rotation axis lies on the equator and is perpendicular to the normal vector, we can write down the rotation axis $\bm{\hat{k}}$ and angle $\beta$ using components of $\bm{\hat{n}}$, so that
\begin{align}
    \bm{\hat{k}} &= \frac{(n_2, -n_1, 0)^{\rm T}}{\sqrt{n_1^2 + n_2^2}}, \\
    \beta &= \arccos(n_3).
\end{align}
We note that, by defining $\bm{g_1}$ to be always on the left of $\bm{g_2}$, there is no ambiguity in the definition of the rotation axis, and the rotation angle will always lie between $[0, \frac{\pi}{2}]$.
Rodrigues' rotation formula \citep{cheng/gupta:1989} then allows us to rotate every point on the sky to the desired frame where the filament pair now lies on the equator, such that for each source galaxy at position $\bm{g_s}$, the new position is at:
\begin{align}
    \bm{g_{\rm new}} = \bm{g_s}\cos\beta+\sin\beta(\bm{\hat{k}}\times\bm{g_s})+(\bm{\hat{k}}\cdot\bm{g_s})(1 - \cos\beta)\bm{\hat{k}}
\end{align}
It is worth noting that, because the shear was measured in each galaxy's local (RA, Dec) coordinate frame, the angle of rotation is different for different source galaxies. 
The transformed shear map ($\widetilde{e_1}, \widetilde{e_2}$) for each galaxy is thus given by
\begin{align}
\begin{pmatrix}
\widetilde{e_1} \\
\widetilde{e_2} \\
\end{pmatrix}
=
\begin{pmatrix}
\cos2\phi_s & \sin2\phi_s \\
-\sin2\phi_s & \cos2\phi_s \\
\end{pmatrix}
\begin{pmatrix}
e_1 \\
e_2 \\
\end{pmatrix} \, .
\label{e_rotation}
\end{align}
where for each source galaxy, the rotation angle $\phi_s$ is defined by the angular change of its local coordinate frame, e.g., the declination to the $y$-axis of the filament.
\begin{figure}
    \centering
    \includegraphics[width=0.6\columnwidth]{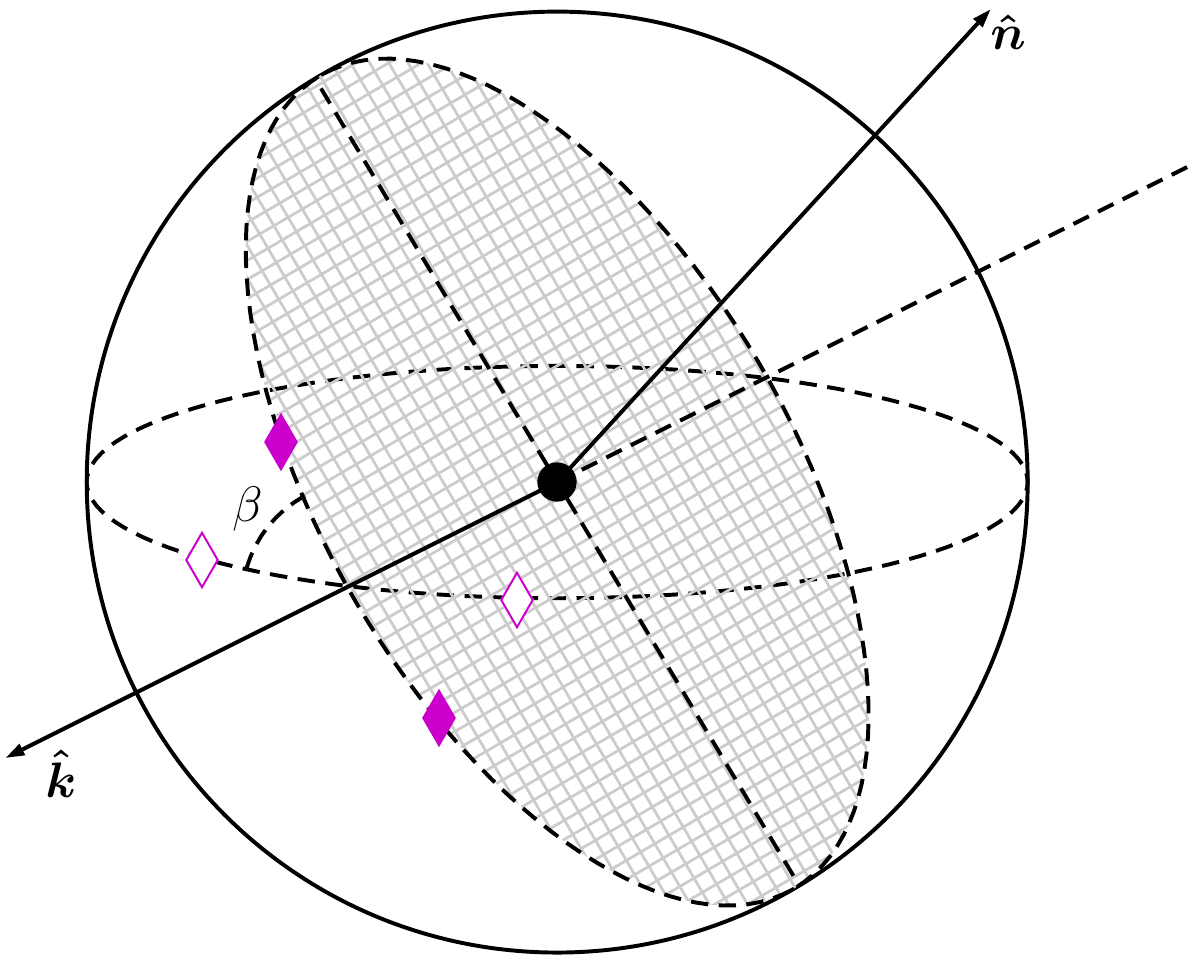}
    \caption{Illustration of the spherical rotation for filament pairs (solid pink diamonds). Open symbols are the rotated positions.}
    \label{fig:sph_rot}
\end{figure}  

\label{lastpage}

\end{document}